# Exploring diamond-like lattice thermal conductivity crystals via feature-based transfer learning

Shenghong Ju,[1,2] Ryo Yoshida,[2,3] Chang Liu,[2,3] Kenta Hongo,[2,4,5] Terumasa Tadano,[2,6] and Junichiro Shiomi,[1,2,7,8,*]

[1]Department of Mechanical Engineering, The University of Tokyo, 7-3-1 Hongo, Bunkyo, Tokyo 113-8656, Japan

[2]Center for Materials Research by Information Integration (CMI2), Research and Services Division of Materials Data and Integrated System (MaDIS), National Institute for Materials Science (NIMS), 1-2-1 Sengen, Tsukuba, Ibaraki 305-0047, Japan

[3]Research Organization of Information and Systems, The Institute of Statistical Mathematics (ISM), 10-3 Midori-cho, Tachikawa, Tokyo 190-8562, Japan

[4]Japan Advanced Institute of Science and Technology (JAIST), 1-1 Asahidai, Nomi, Ishikawa 923-1292, Japan

[5]PRESTO, Japan Science and Technology Agency (JST), Kawaguchi, Saitama 332-0012, Japan

[6]International Center for Young Scientists (ICYS), National Institute for Materials Science, 1-2-1 Sengen, Tsukuba, Ibaraki 305-0047, Japan

[7]Core Research for Evolutional Science and Technology (CREST), Japan Science and Technology Agency (JST), 4-1-8, Kawaguchi, Saitama 332-0012, Japan

[8]RIKEN Center for Advanced Intelligence Project, Nihombashi, Chuo-ku, Tokyo, Japan

**ABSTRACT:** Ultrahigh lattice thermal conductivity materials hold great importance since they play a critical role in the thermal management of electronic and optical devices. Models using machine learning can search for materials with outstanding higher-order properties like thermal conductivity. However, the lack of sufficient data to train a model is a serious hurdle. Herein we show that big data can complement small data for accurate predictions when lower-order feature properties available in big data are selected properly and applied to transfer learning. The connection between the crystal information and thermal conductivity is directly built with a neural network by transferring descriptors acquired through a pre-trained model for the feature property. Successful transfer learning shows the ability of extrapolative prediction, and by screening over 60000 compounds we identify novel crystals that can serve as alternatives to diamond. Even though most materials in the top list are superhard materials, we reveal that superhard property gives high elastic constants and group velocity of phonons in the linear dispersion regime, but it does not necessarily lead to high lattice thermal conductivity because it is determined also by other important factor such as the phonon relaxation time. What's more, the average or maximum dipole polarizability and the van der Waals radius are revealed to be the leading descriptors among those that can also be qualitatively related to anharmonicity.

## ■ INTRODUCTION

The power densities of microelectronic devices and their components continually increase due to advances in the fabrication and integration of advanced materials and structures. Hence, the large thermal density must be quickly removed to guarantee reliable performance. Material innovations in heat spreaders/sinks and thermal interface materials are at the core of the thermal-management challenge. A key element to such innovations is materials with a high lattice thermal conductivity ($\kappa_L$) either as bulk crystals or fillers for composites.

Although metals are generally suitable thermal conductors, insulators have the highest thermal conductivities. In many thermal management applications involving heat spreaders/sinks, electrical insulation is necessary to avoid electric current leakage. Diamond, which has a thermal conductivity of about 2000 $Wm^{-1}K^{-1}$ at room temperature, is a representative bulk material. It is widely used as heat spreaders/sinks for laser diodes and power electronics in the form of bulk or composites to prevent overheating. One drawback is that it sustains thermal damage via oxidation or graphitization at high temperature, significantly altering the thermal properties of the heat spreader/sink.

Cubic and hexagonal boron nitrides have been investigated as alternative materials. Considering the surface affinities with various other materials for composite syntheses and integration, other alternatives should be useful. Although physics insights suggest that some materials exhibit a fairly high $\kappa_L$ such as SiC, BeO, BP, AlN, BeS, GaN, Si, AlP, and GaP, materials with $\kappa_L$ approaching or exceeding 1000 $Wm^{-1}K^{-1}$ are rare.

Single crystal compounds are obvious candidates as alternative high-$\kappa_L$ materials to diamond. However, only a few materials have quantified thermal conductivity values due to difficulties synthesizing single crystals that can be measured in a standardized fashion. Moreover, a material search is extremely cumbersome. Herein we propose utilizing computational techniques to efficiently search for high $\kappa_L$ materials. In recent decades, the development of lattice dynamics methods using interatomic force constants obtained from density functional theories has enabled first-principles calculations of the $\kappa_L$. Simultaneously, databases containing tens of thousands of crystal compounds have been constructed. Examples include Materials Project [1], AFLOW [2], ICSD [3], and AtomWork [4]. However, performing first-principle calculations for all the crystals in the databases is extremely time consuming and unrealistic.

Another option is high throughput screening [5] based on machine learning. The high throughput screening can speed up the discovery of new materials, has been applied in many fields such as catalysis, battery technologies, thermoelectric materials, chemical probes, polymers, and magnetic materials. Motivated by realizing high-performance thermoelectric materials, efforts to apply $\kappa_L$ to crystals have centered on screening ultralow $\kappa_L$ crystals [6-8]. Carrete et al. [6] screened 79000 half-Heusler compounds and found that materials with large atomic radii elements have a lower $\kappa_L$. Seko et al. [7] screened 54779 crystals based on the Gaussian process regression and reported 221 materials with a low $\kappa_L$. Roekeghem et al. [8] extended the screening of mechanically stable compounds at high temperatures using finite-temperature phonon calculations.

One challenge when screening crystals with high/low thermal conductivities is the large gap between the "big data" required for credible machine learning and the "small data" currently available. Although bridging this gap is a general problem in materials informatics, it is especially intense when searching for materials with a preferred thermal conductivity because it involves both the harmonic phonon properties, which are fairly easy to calculate, and anharmonic properties, which are much more expensive to evaluate. As shown in **Figure 1**, "big data" are available for the harmonic phonon property of the three-phonon scattering phase space but not for the thermal conductivity (**Figure 1**). Only "small data" are available for the thermal conductivity due to the heavy calculation required for the anharmonic phonon property.

In this work, we develop a feature-based transfer learning method to overcome the gap. This method begins with a broad search over the entire structure database of crystal compounds for the feature harmonic property, which should be correlated with thermal conductivity. Subsequently, a focused search of the selected candidates is conducted for a high thermal conductivity. Here, we choose the scattering phase space ($P_3$) of the three-phonon scattering process as the feature property because this can be quickly extracted from the harmonic calculation.

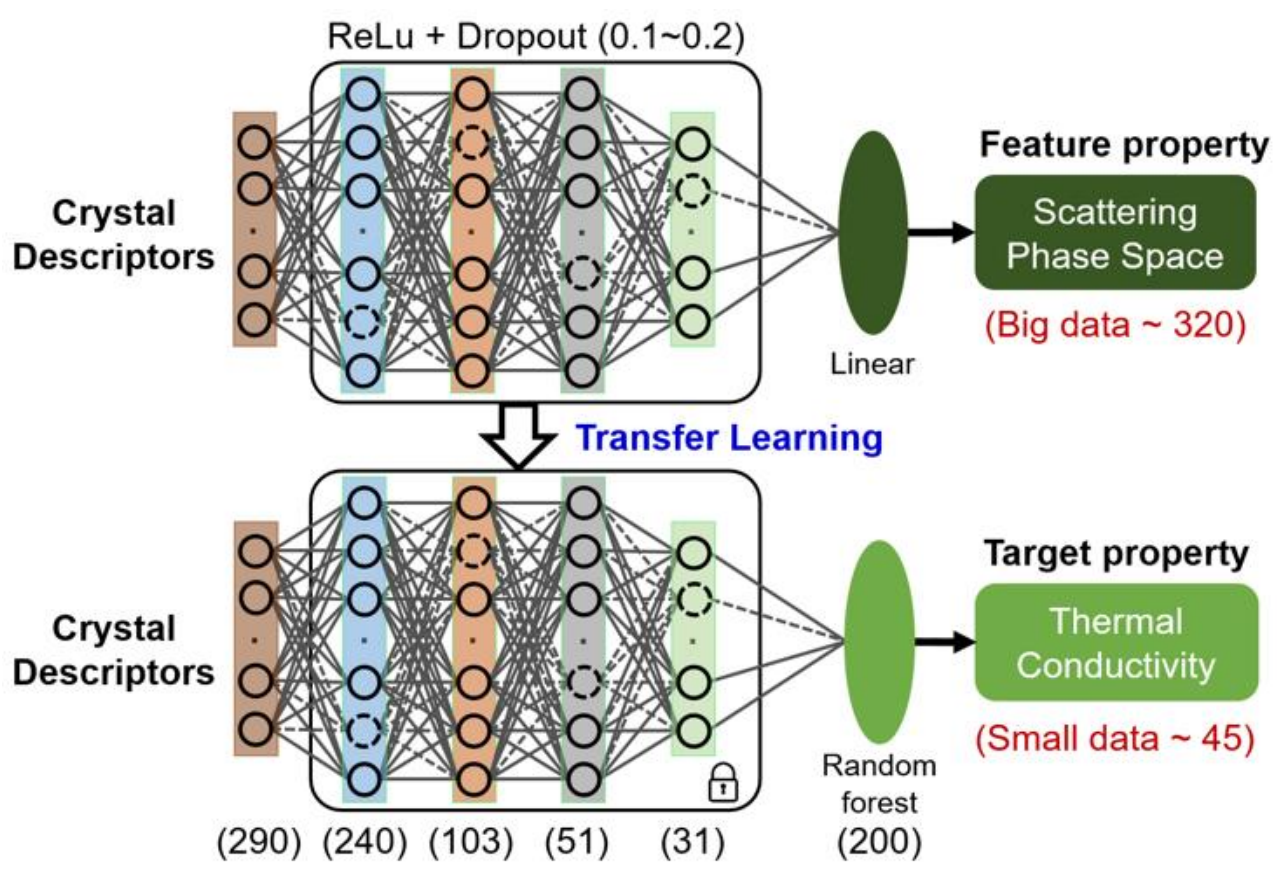

**Figure 1**. **Schematics of feature-based transfer learning**. The transfer learning bridges "big data" (harmonic three-phonon scattering phase space of 320 crystals) and "small data" (thermal conductivity of 45 crystals) to search for ultrahigh lattice thermal conductivity crystals. All neurons (circles) are activated by ReLu (Rectified Linear Unit). Dropout (dash circles and lines) range (0.1 or 0.2) in each hidden layer is randomly chosen. Numbers at the bottom indicate the number of neurons or trees used in each layer of the neural network and random forest model.

## ■ METHODS

**Anharmonic lattice dynamics.** The current approach was formulated on the basis of anharmonic lattice dynamics calculations using the interatomic force constants obtained by first-principles, which are well described elsewhere [9]. The $\kappa_L$ was calculated by solving the Boltzmann transport equation (see Supplementary **Note A**)

$$\boldsymbol{v}_{qs}\cdot\nabla T\left(\frac{\partial n_{qs}}{\partial T}\right)+\left.\frac{\partial n_{qs}}{\partial t}\right|_{scattering}=0, \tag{1}$$

where $n$ is the phonon distribution function, $\boldsymbol{qs}$ is the phonon mode, and $\boldsymbol{v}$ is the group velocity. Although the single-mode relaxation time approximation (see Supplementary **Note B**) is often used to calculate $\kappa_L$, it significantly underestimates $\kappa_L$ for high thermal conductivity crystals. Thus, Eq. (1) must be solved iteratively or directly (see Supplementary **Figure S2** and **Note C**).

As discussed above, a key feature in the current process is $P_3$, which quantifies the phonon scattering channels. The total $P_3$ is calculated as

$$P_3=\frac{1}{N_q}\ _{\boldsymbol{qs}}\ \frac{1}{3m^3}\left(2P_3^{(+)}(\boldsymbol{qs})+P_3^{(\ )}(\boldsymbol{qs})\right), \tag{2}$$

where $m$ is the number of phonon branches and

$$P_3^{(\pm)}(\boldsymbol{qs})=\frac{1}{N_q}\ _{\boldsymbol{q's',q''s''}}\left(\ _{\boldsymbol{qs}}\pm\ _{\boldsymbol{q's'}}\ \ _{\boldsymbol{q''s''}}\right)\ _{\boldsymbol{q\pm q',q''+G}}. \tag{3}$$

Equation (3) indicates that $P_3$ can be calculated solely from harmonic interatomic force constants.

**$P_3$ and $\kappa_L$ data collections.** The candidates were inorganic crystals in the Materials Project [1] database, which currently includes over 60000 entries. Because this study focused on $\kappa_L$, materials with a band gap smaller than 0.1 eV, molecular crys-tals such as $O_2$, $H_2$, $H_2O$, $H_2O_2$, and crystals with hydrogen atoms were excluded. We collected the atom displacement from the phonon database (http://phonondb.mtl.kyoto-u.ac.jp/) and calculated force data of 320 crystals by first-principles (see Supplementary **Table S1**). The ALAMODE package [10] was then used to fit the harmonic interatomic force constants and calculate the $P_3$ values. Besides the $P_3$ data, we also collected thermal conductivity data for 45 materials (see Supplementary **Table S2**).

**Transfer learning.** Transfer learning was performed via a self-developed open-source XenonPy Python package (https://github.com/yoshida-lab/XenonPy). XenonPy calculated 290 compositional features for a given chemical composition using information about the 58 elemental level property data. We initially pre-trained a fully-connected pyramid neural network using 320 instances on $P_3$ and 290 compositional descriptors. All neurons were activated by ReLu (Rectified Linear Unit), and the linear activation function was placed on the output layer, which defines the transformation from the 10 neurons in the last hidden layer to the $P_3$. We produced 1000 pre-trained models on the $P_3$ with randomly generated network structure; number of hidden layers, which ranged between 4 and 6; number of neurons in each layer; and the drop range, which was either 0.1 or 0.2. Subsequently, the best model on the $P_3$ was selected during the 10-fold cross-validation looped within the 320 instances. Except for the output layer, the subnetwork of the selected model was used as both a feature extractor and an input descriptor in the prediction model of the $\kappa_L$. Finally, the random forest (number of trees = 200) model was selected and trained using 45 instances (see Supplementary **Table S2**) of the $\kappa_L$ and the 10-dimensional descriptors acquired through the pre-training process.

## ■ RESULTS

**Performance of transfer learning.** For the given target property ($\kappa_L$), which has limited training data, models on the proxy feature property ($P_3$) are pre-trained using sufficient data to capture the features relevant to the commonality between $\kappa_L$ and $P_3$. Re-purposing such machine learning acquired features on the target task can realize an outstanding prediction ability even with an exceedingly small amount of data. The present study focuses on a specific type of transfer learning using artificial neural networks (**Figure 1**).

The prediction model connecting the crystal structures and $P_3$ was initially trained based on 320 collected instances of $P_3$ (see Supplementary **Table S1**) and 290 compositional descriptors. The subnetwork of the pre-trained model was transferred to train the model connecting the crystal structures and $\kappa_L$ by replacing the linear output layer with the random forest model using 45 $\kappa_L$ data (see Supplementary **Table S2**). Training and validation using the pre-trained and transferred models with optimal hyperparameters validate the learning and prediction (**Figure 2(A)–(B)**).

Using the machine learning models for the $\kappa_L$ and $P_3$ in the high throughput screening with ~60000 crystals in Materials Project [1], we identified the top-14 crystals with the smallest $P_3$ (see Supplementary **Table S3**) from the top-100 prediction list whose $\kappa_L$ were validated using the first-principles based anharmonic lattice dynamics calculations (**Figure 3** and **Table 1**). The calculations details can be seen in Supplementary **Table S7** and **Note D**. The transferred model successfully predicts the 14 crystals even though their $\kappa_L$ lie in the ultrahigh region of 1000–3000 $Wm^{-1}K^{-1}$ (**Figure 2(C)**). It should be noted that the $\kappa_L$ of the 45 training crystals reside in the region smaller than 370 $Wm^{-1}K^{-1}$, which is much lower than the prediction (**Figure 2 (D)**). This indicates that the transferred model possesses "extrapolative prediction". In general, ordinary machine learning is "interpolative", and its prediction ability is applicable only in a neighboring region of the given training instances.

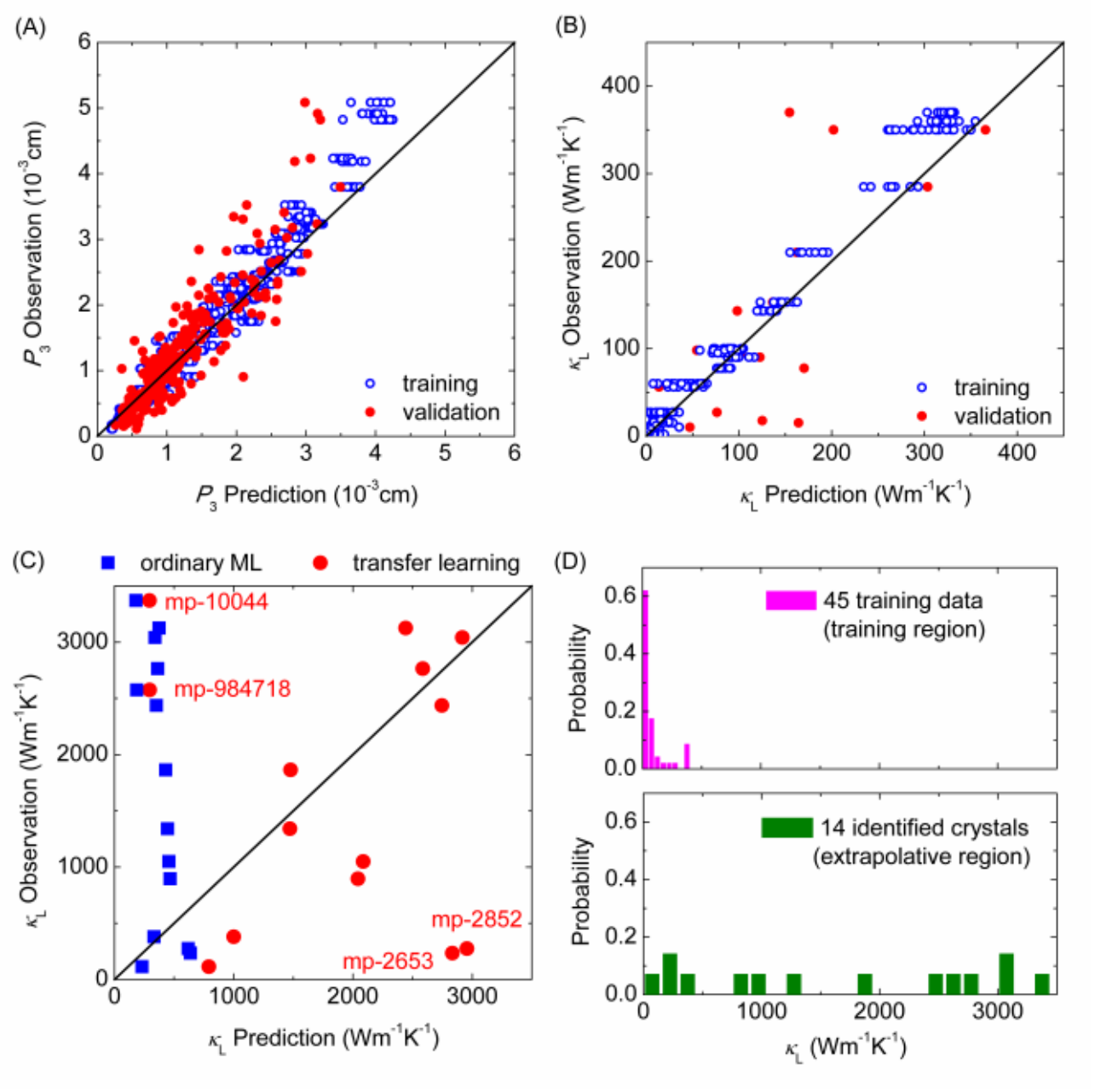

**Figure 2**. **Performance of transfer learning**. Training and validation for (A) the pre-trained $P_3$ model (mean absolute error = 0.000237 cm) and (B) transferred $\kappa_L$ model (mean absolute error = 30.8528 $Wm^{-1}K^{-1}$). Hollow and solid dots denote the results of training and testing in the ten-fold cross-validation for the best prediction model. (C) Comparison between ordinary machine learning and transfer learning. (D) $\kappa_L$ distribution of 45 training and 14 identified crystals.

Indeed, a neural network directly trained using the 45 samples performs rather poorly for the 14 crystals because the predicted $\kappa_L$ never exceeded 600 $Wm^{-1}K^{-1}$ (**Figure 2 (C)** and see Supplementary **Table S4**). The pre-training process using the 320 instances on $P_3$ should contribute to the acquisition of the extrapolative ability. The pre-trained neural network on $P_3$ is able to represent material structures that are applicable to a broader input space than the one spanned by much fewer instances on $\kappa_L$. This is because the 320 source $P_3$ data contain instances that account for the structure-property relationships relevant to ultrahigh thermal conductivity.

**Top high thermal conductivity crystals.** We identified the top-14 materials by feature-based transfer learning. They are comprised of boron arsenides (BAs), carbon (C), beryllium carbide ($Be_2C$), boron nitride (BN), heterodiamond ($BC_2N$), carbon nitride ($C_3N_4$), and ternary $BeCN_2$. They all have thermal conductivities above 100 $Wm^{-1}K^{-1}$. In fact, 10 exceed 500 $Wm^{-1}K^{-1}$. The top two crystals with the lowest calculated $P_3$ are cubic and wurtzite BAs, which have thermal conductivity over 1000 $Wm^{-1}K^{-1}$.

**Table 1. Top-14 materials with the lowest $P_3$ values and their thermal conductivities calculated by the iterative Boltzmann transport equation solution. xx, yy, and zz indicate the lattice directions.**

| Name | Structure | $P_3$ ($10^{-4}$ cm) | Thermal conductivity ($Wm^{-1}K^{-1}$) | | | | |
|---|---|---|---|---|---|---|---|
| | | | This work | | | Cal. Ref. | Exp. Ref. |
| | | | xx | yy | zz | xx/yy (zz) | xx/yy (zz) |
| cubic BAs | F-43m | 0.6397 | 3411 | 3411 | 3411 | 3170 [11] | 351 [12] |
| wurtzite BAs | P6$_3$mc | 0.9064 | 2947 | 2947 | 1881 | 2380 (1210) [13] | |
| diamond | Fd-3m | 1.0005 | 3048 | 3048 | 3048 | 3450 [11] | 3000 [14] |
| lonsdaleite | P6$_3$/mmc | 1.0335 | 2533 | 2533 | 2122 | 1500 (1270) [15] | |
| C (611426) | P6$_3$/mmc | 1.2437 | 2842 | 2842 | 2675 | | |
| C (616440) | P6$_3$/mmc | 1.2569 | 2583 | 2583 | 4214 | | |
| $Be_2C$ | Fm-3m | 1.2596 | 117 | 117 | 117 | | |
| cubic BN | F-43m | 1.3300 | 1876 | 1876 | 1876 | 1800 [13] | 768 [16] |
| $BC_2N$ (30148) | P222$_1$ | 1.3670 | 895 | 910 | 804 | | |
| wurtzite BN | P6$_3$mc | 1.4394 | 1359 | 1359 | 1305 | 1230 (1040) [13] | |
| cubic $C_3N_4$ | I-43d | 1.4490 | 234 | 234 | 234 | | |
| pseudo $C_3N_4$ | P-43m | 1.4529 | 275 | 275 | 275 | | |
| $BC_2N$ (629458) | Pmm2 | 1.5127 | 1392 | 972 | 784 | | |
| $BeCN_2$ | I-42d | 1.5472 | 351 | 351 | 440 | | |

Recently, many studies have investigated cubic BAs due to their high predicted thermal conductivity (3170 $Wm^{-1}K^{-1}$) [11], which is comparable with diamond. However, these initial experiments measured the thermal conductivity of BAs crystals around 186–350 $Wm^{-1}K^{-1}$ [12, 17, 18]. This difference is attributed mainly to difficulties fabricating single crystals of boron-related materials as well as the complicated synthesis due to the high volatility and toxicity of arsenide atoms. Although the four-phonon scattering process is important for high thermal conductivity materials at high temperatures [19], the theoretical thermal conductivity of BAs remains as high as 2000 $Wm^{-1}K^{-1}$. In fact, recent experiments realized cubic BAs with a thermal conductivity as high as 1000 $Wm^{-1}K^{-1}$ [20-22].

Cubic BAs should be a reasonable demonstration of the effectiveness of the current screen. In this paper, we focus on the thermal conductivity at room temperature. Because the three-phonon scattering rate is much higher than the four-phonon scattering rate, employing three-phonon $P_3$ is reasonable to find high thermal conductivity materials.

Another interesting feature of the top list is that it contains allotropes of known high thermal conductivity cubic materials (diamond, BN, and BAs). These include lonsdaleite, hexagonal diamonds, wurtzite BN, and wurtzite BAs, which have not been studied previously in terms of thermal conductivity. These materials are potential alternatives to their cubic counterparts. Although both cubic BN and wurtzite BN can both be formed by compressing hexagonal BN, wurtzite BN is formed at much lower temperatures (around 2000 K) than cubic ones (3000–4000 K) [23]. Recently, a single-phase wurtzite BN bulk crystal was synthesized directly from a hexagonal BN bulk crystal under 10 GPa and 850 °C [24]. If the wurtzite structure has similar merits in other species, then BAs may be the first to benefit.

The list of top-100 materials with the smallest $P_3$ also contains some typical high-thermal-conductivity materials like SiC (mp-ID: 8062), GaN (mp-ID: 830), and AlN (mp-ID: 1700) (see Supplementary **Table S3**). Additionally, the top-100 list includes different crystal structures of GaN (mp-ID: 804) and AlN (mp-ID: 1330), which may display high thermal conductivities. Layered structure materials such as hexagonal BN and graphite have high thermal conductivities in the in-plane direction, but their out-of-plane thermal conductivity is very low due to the weak atomistic interaction. Hexagonal BN (mp-ID: 984) is in the top-100 prediction list as well as other BNs with layered structures. Examples include mp-ID: 7991, 685145, 13150, 604884, 629015, and 569655. Moreover, the list has some graphite structures (mp-ID: 568806, 632329, 990448, 568286 and 569304). In this work, we evaluate the thermal conductivity with the scattering phase space, which is a scalar parameter that tends to suggest crystals with a high thermal conductivity in all three-lattice directions. The experimental conditions are the main reason why 2D materials such as hexagonal BN and graphite do not appear in our top-14 list.

**Hardness versus thermal conductivity.** An interesting feature of the screening results is that most of the top-14 list are superhard materials, including diamond, carbon nitride, born nitride, and heterodiamond. The Vickers hardness of diamond is around 115 GPa [25], which is the highest among reported superhard crystals, including cubic BN (62 GPa) [26], cubic $BC_2N$ (76 GPa) [25], $C_3N_4$ (37–90 GPa) [27], and $BeCN_2$ (37 GPa ) [28]. $Be_2C$ has a Knoop hardness of 2410 kg $mm^{-2}$, whereas diamond has a Knoop hardness of 7000 kg $mm^{-2}$ [29].

The shear modulus is roughly proportional to the hardness. In the past two decades, it has been used as a guide for theoretical predictions of hard materials. **Figure 4 (A)** plots the calculated shear modulus versus the average $\kappa_L$. Materials with a superhard property do not necessarily result in a high $\kappa_L$. The shear modulus of cubic and wurtzite BAs is only around 123 GPa which is the lowest among the top-14 materials, but their thermal conductivities exceed 2500 $Wm^{-1}K^{-1}$. Another example is superhard cubic silicon nitride ($Si_3N_4$). Although its reported Vickers hardness is around 35 GPa [30], the calculated thermal conductivity is only around 81 $Wm^{-1}K^{-1}$ [31], which is much lower than that of $BeCN_2$ with the same order of hardness.

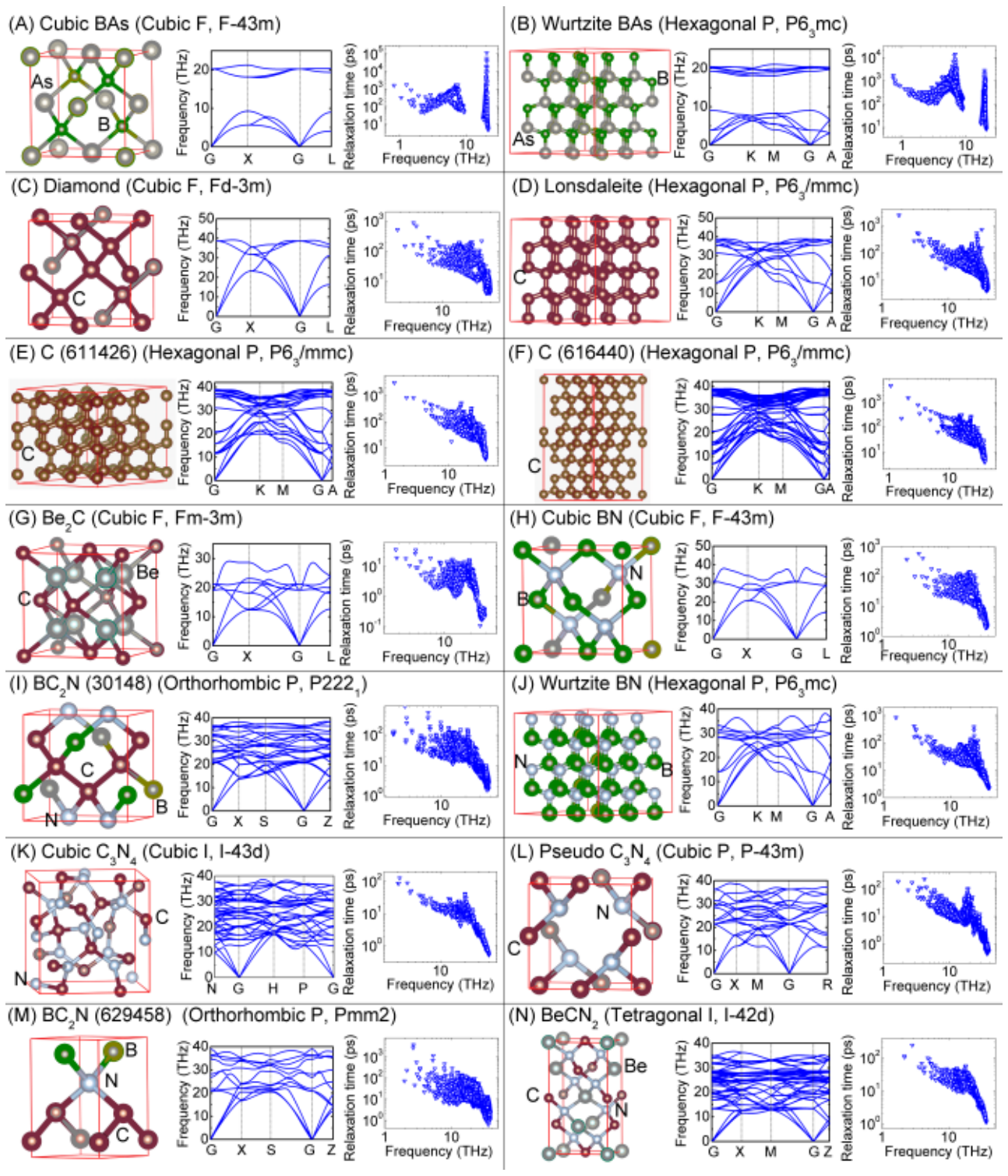

**Figure 3. Top-14 crystals with the lowest calculated $P_3$.** Crystal structures, phonon dispersions, and phonon relaxation times of the top-14 materials with the lowest calculated $P_3$. Parenthesis () indicates the crystal structure.

**Figures 4 (B)-(D)** show the average group velocity, heat capacity, and phonon relaxation time of the top-14 materials, respectively. Cubic and wurtzite BAs have group velocities lower than other hard materials, but they have the highest relaxation times. Consequently, they have a high $\kappa_L$ comparable with diamond. Although superhardness means large elastic constants and group velocities of phonons in the linear dispersion regime, $\kappa_L$ is also determined by other factors such as the phonon relaxation time (i.e., $P_3$ and anharmonic scattering amplitude). In addition, the properties of phonons with nonlinear dispersions largely influence the $\kappa_L$. This highlights the necessity and importance of developing rapid screening models to explore high $\kappa_L$ crystals from databases.

Superhard materials consisting of B, C, and N atoms like BN, $BC_2N$, and $C_3N_4$ are advantageous over diamond in terms of stability and oxidation because the covalent bond energies between B-N (–117.19 eV) and C-N (–141.74 eV) are stronger than that of C-C (–103.64 eV) [32]. Mixing diamond with BN as a starting material may create new BCN alloy compounds under high pressure and temperature, which are more stable thermally and chemically than diamond and harder than BN.

Among ternary BCN compounds, heterodiamond in the form of $BC_2N$ has gained some attention. Heterodiamond has various structural forms ranging from layered graphite-like and diamond structures. However, the top-14 list includes two cubic $BC_2N$ (mp-ID: 30148 and 629458) structures with thermal conductivities around 784–1392 $Wm^{-1}K^{-1}$. Polycrystalline cubic $BC_2N$ materials have been synthesized from hexagonal BN at 20 GPa and 2200–2250 K [33] and from graphite-like $BC_2N$ above 18 GPa and 2100–2200 K [25, 26]. The measured hardness of synthesized cubic $BC_2N$ is higher than that of a cubic BN single crystal but lower than diamond [25, 26]. The current finding that cubic $BC_2N$ has a high potential to be an ultrahigh thermal conductor is motivation to improve the synthesis techniques of heterodiamond.

$C_3N_4$ is another interesting superhard material with a reported hardness around 37–90 GPa [27]. Here we found two cubic structures of $C_3N_4$ in the top-14 list. One is pseudo $C_3N_4$ (mp-ID: 571653) and the other is cubic $C_3N_4$ (mp-ID: 2852) (Fig. 3). The former has a defect zincblende structure with a hole in the central region of the unit cell. The latter has as many as 14 atoms in the primitive unit cell (see Supplementary **Figure S3**), which gives rise to complex phonon modes with many branches. Despite their apparent defective and complex structures, their thermal conductivities exceed 200 $Wm^{-1}K^{-1}$. Efforts have been made to synthesize different phases of carbon nitrides [34]. Martin-Gil *et al*. [34] synthesized a pseudo $C_3N_4$ by a chemical precursor route under 800 °C and claims the process is scalable.

Lonsdaleite, which is also a wurtzite structure, has been studied previously as its hardness is comparable or even harder than diamond [35]. Its thermal conductivity (2122–2533 $Wm^{-1}K^{-1}$) is also comparable with diamond. Recently, polycrystalline lonsdaleite has been successfully synthesized in a diamond anvil cell at 100 GPa and 400 °C [36] using graphitic layers, which provide a low energy barrier for progressive transformation from graphite to lonsdaleite. The synthesis temperatures are well below those previously reported for lonsdaleite [35].

The other two hexagonal diamonds (mp-ID: 611426 and 616440) combine the structure features of diamond and lonsdaleite (see Supplementary **Figure S4**). These can be described as diamond/lonsdaleite superlattices. Their thermal conductivities are comparable with diamond.

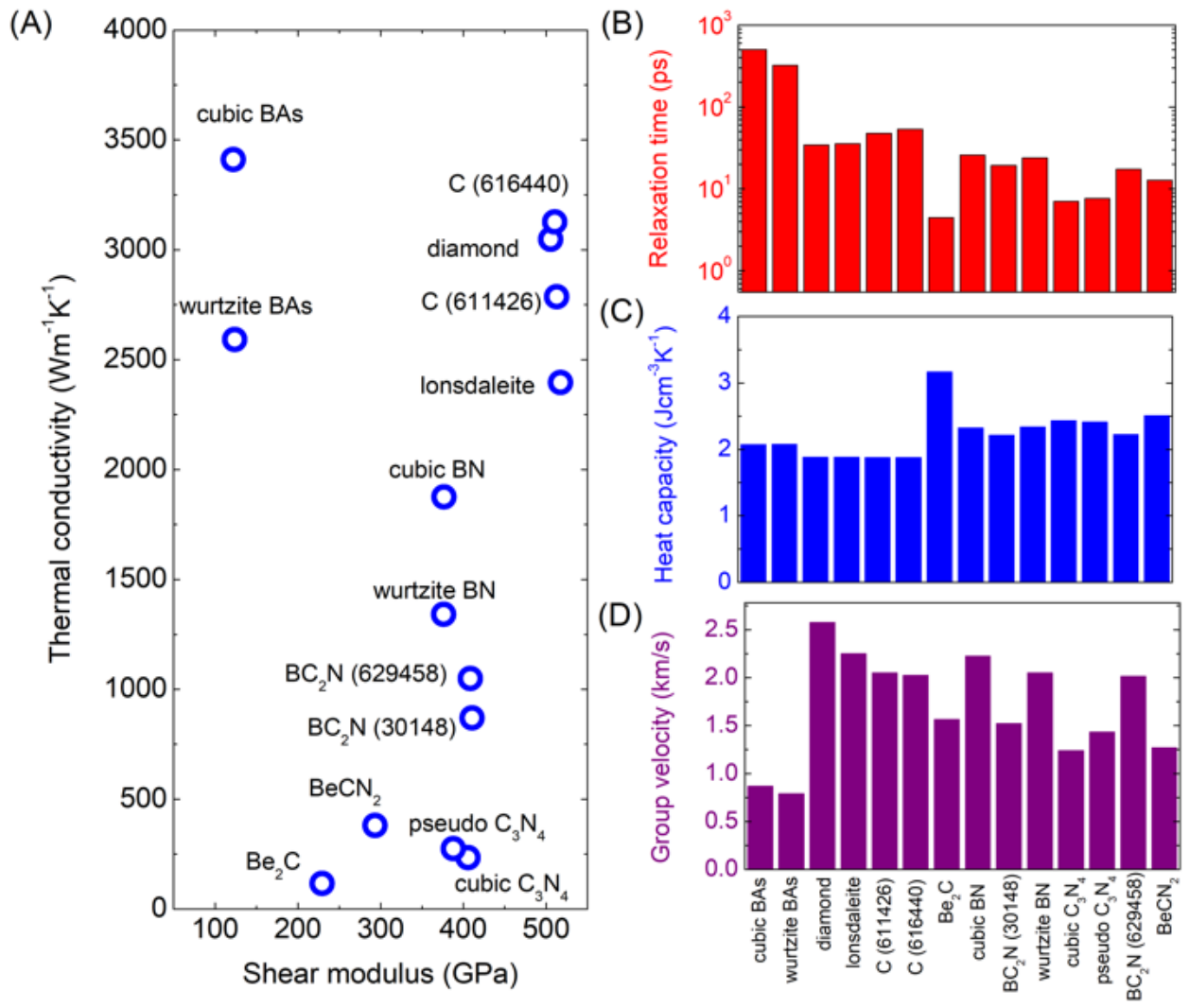

**Figure 4**. **Hardness versus thermal conductivity and comparison of parameters related with thermal conductivity**. (A) Thermal conductivity versus shear modulus for the top-14 materials. (B)-(D) Average group velocity, heat capacity, and relaxation time of the top-14 materials.

**Knowledge gained from transfer learning**. The comparison between the pre-trained and transferred models provides some important physical indications. Here, the pre-trained model only involves the harmonic phonon properties, whereas the transferred model also requires anharmonic properties (i.e., the magnitude of the three-phonon scattering obtained from cubic interatomic force constants). The success of transfer learning means that the correlation between $P_3$ and $\kappa_L$ can be learned from that between the basic crystal structure information and $P_3$, revealing an underlying commonality of the descriptors corresponding to the harmonic and anharmonic properties.

To understand the differences in how to recognize structure-property relationships for the pre-trained and transferred models, we created a descriptor-property heatmap (**Figure 5**). For each model, the 290 compositional descriptors of the ~60000 candidates to be screened are displayed onto the heatmap. The candidate materials in the heatmap are sorted according to a descending order of the predicted values on $P_3$ or $\kappa_L$. Such a visualization reveals the presence of key descriptors relevant to the pattern recognition inherent in the trained model. Irrelevant or relevant descriptors might exhibit completely random or non-random patterns such as a linear trend along with the ordered predicted properties from top to bottom.

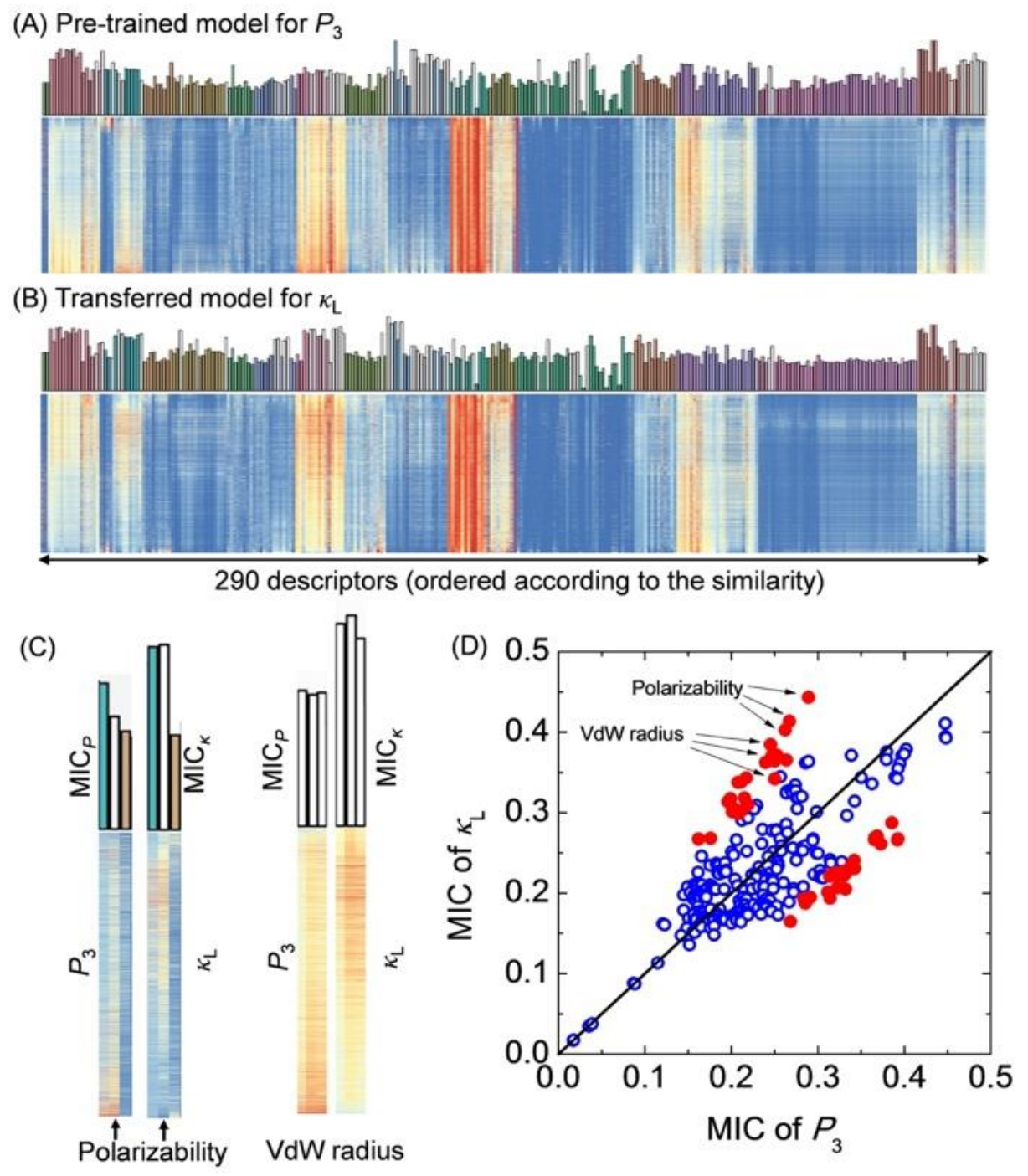

**Figure 5**. **Descriptor-property heatmap and maximal information coefficient (MIC) scores**. Descriptor-property heatmap of (A) the pre-trained model for $P_3$ and (B) transferred model for $\kappa_L$. Top bar plots show the MIC scores for each descriptor. (C) Enlarged heatmap for the descriptors of polarizability and the van der Waals (VdW) radius. (D) Distribution of the MIC scores for the 290 descriptors with respect to $P_3$ and $\kappa_L$. Solid red dots denote 57 key descriptors with the MIC scores exhibiting significant differences between $P_3$ and $\kappa_L$ (see Supplementary **Table S5** and **S6**).

We investigated the underlying mechanisms responsible for the successful transfer from $P_3$ to $\kappa_L$. We aimed to interpret key features that distinguish between the pre-trained and post-transferred models. To quantitatively assess the dependency (relevance or association) between each descriptor and the predicted $P_3$ or $\kappa_L$, we introduced the maximal information coefficient (MIC). MIC is a common measure of nonlinear correlations in bivariate random variables [37].

The bar plots in **Figure 5** show the MIC scores of the 290 descriptors with regard to $P_3$ and $\kappa_L$. While most of the realized MICs do not change significantly between $P_3$ and $\kappa_L$, some descriptors exhibit either a significant increase or decrease in the MICs. By looking at the descriptors discriminating the mechanisms regulating $P_3$ and $\kappa_L$ where the differences of $MIC_P$ - $MIC_\kappa$ are ± 0.09 (see Supplementary **Tables S5** and **S6**), we can identify descriptors that determine the $P_3$ and $\kappa_L$.

The descriptors related with the average or maximum dipole polarizability and van der Waals (VdW) radius over all atoms in the crystal compounds are relevant to $\kappa_L$ but not to $P_3$. The polarizability generally correlates with the interactions between electrons and the nucleus. Atoms with a larger number of electrons or atomic radius tend to have a high polarizability. In crystal compounds with a large electronic polarizability, the displacement or force perturbation can be easily transferred and persists over a long range via the orbital electron interaction with the nucleus. A typical example is rocksalt IV–VI materials like PbTe crystal, where the resonant bonding [38] and the corresponding large anharmonic interatomic-force constants are manifested due to the long-range polarization.

The maximum VdW radius that characterizes the non-bonded interactions between atoms also affects the anharmonicity in crystals. The VdW radius is related with the polarizability via the relation $V_w = \alpha/(4\pi\varepsilon_0)$, where $\alpha$ is the polarizability, $\varepsilon_0$ is the relative permittivity, and $V_w$ is the VdW volume, which is given by the VdW radius. Since the descriptors correlated with $\kappa_L$ but not with $P_3$ should govern the linear output layer between the subnetwork pre-trained by $P_3$ and the final $\kappa_L$, it makes sense that polarizability and the VdW radius related to anharmonicity are the descriptors. However, it is meaningful to quantitatively identify that the average or maximum dipole polarizability and the VdW radius are the leading descriptors among those that can also be qualitatively related to anharmonicity. This provides a search direction. In addition to a small $P_3$, the low average or maximum dipole polarizability and VdW radius are necessary to achieve a high $\kappa_L$.

## ■ CONCLUSIONS

In summary, we screened over 60000 crystal compounds with phonon $P_3$ as the feature quantity and identified a set of semiconducting compounds with high thermal conductivities. Screening was performed based on our developed feature-based transfer learning, which bridges the gaps between the "big data" required for credible machine learning and the "small data" of thermal conductivity. Transfer learning directly models the connection between the basic crystal information and the thermal conductivity with a neural network by transferring descriptors acquired through pre-training for $P_3$. The successful prediction of high thermal conductivity crystals demonstrates the advantage of extrapolative prediction via transfer learning, and reveals the descriptors that are dominantly correlated with the anharmonic phonon properties.

The final, obtained materials in the top-14 list by feature-based and transfer learning screening all show high thermal conductivities, including boron arsenides (BAs), carbon (C), boron nitride (BN), and heterodiamond ($BC_2N$). They have thermal conductivities on the order of 1000 $Wm^{-1}K^{-1}$, validating the accuracy and high-efficiency of the developed screening method. Although most of these are superhard materials with a large group velocity, the results are non-trivial. It has been observed that for some superhard materials, a large dispersion could lead to a large $P_3$ and limit thermal conductivity. Because screening via $P_3$ prefers to search crystals with a high thermal conductivity in three different lattice directions, 2D materials such as hexagonal BN and graphite do not appear in our top-14 list due to the weak atomistic interaction in the out-of-plane direction.

The screening also identified known materials that have yet to be studied in the context of heat transport. These include two types of novel carbon crystals with mixed phases of diamond and lonsdaleite and two phases of $BC_2N$. These materials may be advantageous over well-explored high thermal conductivity materials in terms of thermodynamic stability and facileness of synthesis. These findings should contribute to next-generation thermal management technology by broadening the alternatives of high thermal conductivity materials and adding degrees of freedom to their surface affinities with various other materials for composite syntheses and integration.

## ASSOCIATED CONTENT

**Supporting Information**.

The PDF file includes:
**Table S1**. Training data set of $P_3$ for 320 materials.
**Table S2**. Training data set of $\kappa_L$ for 45 materials.
**Table S3**. Top-100 materials with lowest $P_3$.
**Table S4**. Comparison of ordinary machine learning and transfer learning.
**Table S5**. Descriptors relevant to $\kappa_L$ and irrelevant to $P_3$.
**Table S6**. Descriptors relevant to $P_3$ and irrelevant to $\kappa_L$.
**Table S7**. Comparison of relaxation time approximation and iterative Boltzmann transport equation solution for top-14 materials.
**Table S8**. Parameter settings for ShengBTE calculation.
**Note A**. Phonon Boltzmann transport equation.
**Note B**. Relaxation time approximation.
**Note C**. Iterative solution of Boltzmann transport equation.
**Note D.** Density-functional theory calculation details.
**Figure S1**. Types of three-phonon scattering processes.
**Figure S2**. Comparison of relaxation time approximation and iterative Boltzmann transport equation solution.
**Figure S3**. The looks-defective and complex structures of pseudo and cubic $C_3N_4$.
**Figure S4**. Comparison of carbon allotropes.

This material is available free of charge via the Internet at http://pubs.acs.org.

## AUTHOR INFORMATION

### Corresponding Author

* Email: shiomi@photon.t.u-tokyo.ac.jp.

### Author Contributions

S.J. carried out the first-principles calculations and analyzed the results. R.Y. and C.L. performed the transfer learning. S.J., R.Y., and J.S. wrote the manuscript. All the authors discussed the results and implications, and provided comments on the manuscript.

## ACKNOWLEDGMENT

We thank Atsushi Togo for providing the displacement and force data in the Phonon database for the $P_3$ calculation, and Takuma Shiga for the useful discussion about first-principles calculations. This work was supported by the "Materials research by Information Integration" Initiative (MI2I) project, CREST Grant No. JPMJCR16Q5, and PRESTO Grant No. JPMJPR16NA from the Japan Science and Technology Agency (JST), and KAKENHI Grants No. 16H04274, Grants No. 19K14902, and Grants No. JP17K17762 from the Japan Society for the Promotion of Science (JSPS). The calculations in this work were performed using the supercomputer facilities from JAIST, and the Institute for Solid State Physics, the University of Tokyo.

## REFERENCES

1. Jain, A.; Ong, S. P.; Hautier, G.; Chen, W.; Richards, W. D.; Dacek, S.; Cholia, S.; Gunter, D.; Skinner, D.; Ceder, G.; Persson, K. A., Commentary: The Materials Project: A materials genome approach to accelerating materials innovation. Apl Mater 2013, 1, (1), 011002.
2. Curtarolo, S.; Setyawan, W.; Wang, S.; Xue, J.; Yang, K.; Taylor, R. H.; Nelson, L. J.; Hart, G. L. W.; Sanvito, S.; Buongiorno-Nardelli, M.; Mingo, N.; Levy, O., AFLOWLIB.ORG: A distributed materials properties repository from high-throughput ab initio calculations. Computational Materials Science 2012, 58, 227-235.
3. Belsky, A.; Hellenbrandt, M.; Karen, V. L.; Luksch, P., New developments in the Inorganic Crystal Structure Database (ICSD): accessibility in support of materials research and design. Acta Crystallographica Section B 2002, 58, (3 Part 1), 364-369.
4. Xu, Y.; Yamazaki, M.; Villars, P., Inorganic Materials Database for Exploring the Nature of Material. Jpn J Appl Phys 2011, 50, (11), 11RH02.
5. Curtarolo, S.; Hart, G. L.; Nardelli, M. B.; Mingo, N.; Sanvito, S.; Levy, O., The high-throughput highway to computational materials design. Nature Materials 2013, 12, (3), 191-201.
6. Carrete, J.; Li, W.; Mingo, N.; Wang, S.; Curtarolo, S., Finding Unprecedentedly Low-Thermal-Conductivity Half-Heusler Semiconductors via High-Throughput Materials Modeling. Physical Review X 2014, 4, (1), 011019.
7. Seko, A.; Togo, A.; Hayashi, H.; Tsuda, K.; Chaput, L.; Tanaka, I., Prediction of Low-Thermal-Conductivity Compounds with First-Principles Anharmonic Lattice-Dynamics Calculations and Bayesian Optimization. Phys Rev Lett 2015, 115, (20), 205901.
8. Roekeghem, A. v.; Carrete, J.; Oses, C.; Curtarolo, S.; Mingo, N., High-Throughput Computation of Thermal Conductivity of High-Temperature Solid Phases: The Case of Oxide and Fluoride Perovskites. Physical Review X 2016, 6, (4), 041061.
9. Esfarjani, K.; Stokes, H. T., Method to extract anharmonic force constants from first principles calculations. Phys Rev B 2008, 77, (14), 144112.
10. Tadano, T.; Gohda, Y.; Tsuneyuki, S., Anharmonic force constants extracted from first-principles molecular dynamics: applications to heat transfer simulations. J Phys Condens Matter 2014, 26, (22), 225402.
11. Lindsay, L.; Broido, D. A.; Reinecke, T. L., First-principles determination of ultrahigh thermal conductivity of boron arsenide: a competitor for diamond? Phys Rev Lett 2013, 111, (2), 025901.
12. Tian, F.; Song, B.; Lv, B.; Sun, J.; Huyan, S.; Wu, Q.; Mao, J.; Ni, Y.; Ding, Z.; Huberman, S.; Liu, T.-H.; Chen, G.; Chen, S.; Chu, C.-W.; Ren, Z., Seeded growth of boron arsenide single crystals with high thermal conductivity. Appl Phys Lett 2018, 112, (3), 031903.
13. Togo, A.; Chaput, L.; Tanaka, I., Distributions of phonon lifetimes in Brillouin zones. Phys Rev B 2015, 91, (9), 094306.
14. Wei, L.; Kuo, P. K.; Thomas, R. L.; Anthony, T. R.; Banholzer, W. F., Thermal conductivity of isotopically modified single crystal diamond. Phys Rev Lett 1993, 70, (24), 3764-3767.

15. Li, W.; Carrete, J.; A. Katcho, N.; Mingo, N., ShengBTE: A solver of the Boltzmann transport equation for phonons. Computer Physics Communications 2014, 185, (6), 1747-1758.
16. Novikov, N. V.; Osetinskaya, T. D.; Shul'zhenko, A. A.; Podoba, A. P.; Sokolov, A. N.; Petmsha, I. A., Dopov. Akad. Nauk. Ukr. RSR, Ser. A Fiz. Mat. Tekh. Nauki USSR 1983, 72-75.
17. Lv, B.; Lan, Y.; Wang, X.; Zhang, Q.; Hu, Y.; Jacobson, A. J.; Broido, D.; Chen, G.; Ren, Z.; Chu, C.-W., Experimental study of the proposed super-thermal-conductor: BAs. Applied Physics Letters 2015, 106, (7), 074105.
18. Kim, J.; Evans, D. A.; Sellan, D. P.; Williams, O. M.; Ou, E.; Cowley, A. H.; Shi, L., Thermal and thermoelectric transport measurements of an individual boron arsenide microstructure. Appl Phys Lett 2016, 108, (20), 201905.
19. Feng, T.; Lindsay, L.; Ruan, X., Four-phonon scattering significantly reduces intrinsic thermal conductivity of solids. Physical Review B 2017, 96, (16), 161201(R).
20. Tian, F.; Song, B.; Chen, X.; Ravichandran, N. K.; Lv, Y.; Chen, K.; Sullivan, S.; Kim, J.; Zhou, Y.; Liu, T.-H.; Goni, M.; Ding, Z.; Sun, J.; Gamage, G. A. G. U.; Sun, H.; Ziyaee, H.; Huyan, S.; Deng, L.; Zhou, J.; Schmidt, A. J.; Chen, S.; Chu, C.-W.; Huang, P. Y.; Broido, D.; Shi, L.; Chen, G.; Ren, Z., Unusual high thermal conductivity in boron arsenide bulk crystals. Science 2018, 361, 582.
21. Li, S.; Zheng, Q.; Lv, Y.; Liu, X.; Wang, X.; Huang, P. Y.; Cahill, D. G.; Lv, B., High thermal conductivity in cubic boron arsenide crystals. Science 2018, 361, 579.
22. Kang, J. S.; Li, M.; Wu, H.; Nguyen, H.; Hu, Y., Experimental observation of high thermal conductivity in boron arsenide. Science 2018, 361, 575.
23. Vel, L.; Demazeau, G.; Etourneau, J., Cubic boron nitride: synthesis, physicochemical properties and applications. Materials Science and Engineering: B 1991, 10, (2), 149-164.
24. Deura, M.; Kutsukake, K.; Ohno, Y.; Yonenaga, I.; Taniguchi, T., Nanoindentation measurements of a highly oriented wurtzite-type boron nitride bulk crystal. Jpn J Appl Phys 2017, 56, (3), 030301.
25. Solozhenko, V. L.; Andrault, D.; Fiquet, G.; Mezouar, M.; Rubie, D. C., Synthesis of superhard cubic BC2N. Appl Phys Lett 2001, 78, (10), 1385-1387.
26. Solozhenko, V. L.; Dub, S. N.; Novikov, N. V., Mechanical properties of cubic BC2N, a new superhard phase. Diam Relat Mater 2001, 10, (12), 2228-2231.
27. Fan, Q.; Chai, C.; Wei, Q.; Yang, Y., Two Novel C3N4 Phases: Structural, Mechanical and Electronic Properties. Materials 2016, 9, (6), 427.
28. Gou, H.-Y.; Gao, F.-M.; Zhang, J.-W.; Li, Z.-P., Structural transition, dielectric and bonding properties of BeCN2. Chinese Phys B 2011, 20, (1), 016201.
29. Somorjai, G. A.; Li, Y., Introduction to Surface Chemistry and Catalysis, Second Edition. John Wiley & Sons: 2010.
30. Nishiyama, N.; Ishikawa, R.; Ohfuji, H.; Marquardt, H.; Kurnosov, A.; Taniguchi, T.; Kim, B. N.; Yoshida, H.; Masuno, A.; Bednarcik, J.; Kulik, E.; Ikuhara, Y.; Wakai, F.; Irifune, T., Transparent polycrystalline cubic silicon nitride. Sci Rep 2017, 7, 44755.
31. Tatsumi, K.; Togo, A.; Tanaka, I., First-principles calculation of the lattice thermal conductivities of Si3N4. arXiv:1612.08480 2016.
32. Chen, P., Material Science and Engineering: Proceedings of the 3rd Annual 2015 International Conference on Material Science and Engineering (ICMSE2015, Guangzhou, Guangdong, China, 15-17 May 2015). CRC Press: 2016.
33. Zhao, Y.; He, D. W.; Daemen, L. L.; Shen, T. D.; Schwarz, R. B.; Zhu, Y.; Bish, D. L.; Huang, J.; Zhang, J.; Shen, G.; Qian, J.; Zerda, T. W., Superhard B–C–N materials synthesized in nanostructured bulks. Journal of Materials Research 2011, 17, (12), 3139-3145.
34. Martin-Gil, J.; Martin-Gil, F. J.; Sarikaya, M.; Qian, M.; José-Yacamán, M.; Rubio, A., Evidence of a low compressibility carbon nitride with defect-zincblende structure. J Appl Phys 1997, 81, (6), 2555-2559.
35. Bundy, F. P.; Kasper, J. S., Hexagonal Diamond—A New Form of Carbon. The Journal of Chemical Physics 1967, 46, (9), 3437-3446.
36. Shiell, T. B.; McCulloch, D. G.; Bradby, J. E.; Haberl, B.; Boehler, R.; McKenzie, D. R., Nanocrystalline hexagonal diamond formed from glassy carbon. Sci Rep 2016, 6, 37232.
37. Speed, T., A Correlation for the 21st Century. Science 2011, 334, (6062), 1502-1503.
38. Lee, S.; Esfarjani, K.; Luo, T.; Zhou, J.; Tian, Z.; Chen, G., Resonant bonding leads to low lattice thermal conductivity. Nature Communications 2014, 5, 3525.

Supporting Information for

# Exploring diamond-like lattice thermal conductivity crystals via feature-based transfer learning

Shenghong Ju, Ryo Yoshida, Chang Liu, Kenta Hongo, Terumasa Tadano, and Junichiro Shiomi*

Corresponding E-mail: shiomi@phonton.t.u-tokyo.ac.jp

**Table S1. Training data set of $P_3$ for 320 materials**.

| id | Name | $P_3$ (cm) | id | Name | $P_3$ (cm) |
|---|---|---|---|---|---|
| mp-28797 | YHSe | 0.000111 | mp-3992 | PrSF | 0.000953 |
| mp-36248 | $H_4BrN$ | 0.000152 | mp-5760 | NdSF | 0.000955 |
| mp-34337 | $H_4NCl$ | 0.000171 | mp-10931 | HoSF | 0.000957 |
| mp-24012 | HoHSe | 0.000187 | mp-8407 | $Li_3LaP_2$ | 0.000958 |
| mp-1541 | BeSe | 0.000194 | mp-27546 | CaClF | 0.000958 |
| mp-23703 | LiH | 0.000230 | mp-10932 | ErSF | 0.000959 |
| mp-830 | GaN | 0.000233 | mp-11107 | $Ac_2O_3$ | 0.000964 |
| mp-2542 | BeO | 0.000252 | mp-2488 | SiOs | 0.000966 |
| mp-1265 | MgO | 0.000262 | mp-8402 | $RbMgF_3$ | 0.000966 |
| mp-252 | BeTe | 0.000276 | mp-1873 | $ZnF_2$ | 0.000969 |
| mp-7599 | BeO | 0.000298 | mp-10402 | $TiTl_2F_6$ | 0.000970 |
| mp-8062 | SiC | 0.000302 | mp-10250 | $BaLiF_3$ | 0.000971 |
| mp-11917 | $Mg(BeN)_2$ | 0.000309 | mp-8136 | ThSO | 0.000975 |
| mp-1479 | BP | 0.000336 | mp-5878 | $KZnF_3$ | 0.000976 |
| mp-1700 | AlN | 0.000337 | mp-9006 | $Ho_2CF_2$ | 0.000989 |
| mp-2172 | AlAs | 0.000361 | mp-553303 | $CsCu_3O_2$ | 0.000989 |
| mp-2624 | AlSb | 0.000364 | mp-4511 | $La_2SO_2$ | 0.001008 |
| mp-23737 | $KMgH_3$ | 0.000367 | mp-8430 | KLiS | 0.001008 |
| mp-8881 | AlAs | 0.000381 | mp-10615 | BaLiP | 0.001017 |
| mp-23870 | NaH | 0.000394 | mp-7233 | $La_2SeO_2$ | 0.001018 |
| mp-422 | BeS | 0.000397 | mp-5394 | LaSF | 0.001020 |
| mp-8756 | KLiSe | 0.000398 | mp-11147 | $Na_5CuSO_2$ | 0.001023 |
| mp-617 | $PtO_2$ | 0.000404 | mp-3762 | $VCu_3S_4$ | 0.001024 |
| mp-23818 | $BaLiH_3$ | 0.000404 | mp-1415 | CaSe | 0.001027 |
| mp-625548 | $Cd(HO)_2$ | 0.000415 | mp-661 | AlN | 0.001028 |
| mp-24084 | KH | 0.000420 | mp-8278 | $Ba(MgP)_2$ | 0.001030 |
| mp-2490 | GaP | 0.000425 | mp-5663 | $BaCeO_3$ | 0.001031 |
| mp-632667 | $YbH_2$ | 0.000439 | mp-2472 | SrO | 0.001034 |
| mp-2605 | CaO | 0.000449 | mp-1190 | ZnSe | 0.001035 |
| mp-3216 | $Li_2ZrN_2$ | 0.000454 | mp-8579 | $Ba(AgS)_2$ | 0.001036 |
| mp-1138 | LiF | 0.000455 | mp-239 | $BaS_3$ | 0.001045 |

| id | Name | $P_3$ (cm) | id | Name | $P_3$ (cm) |
|---|---|---|---|---|---|
| mp-682 | NaF | 0.000475 | mp-9569 | $Ca(ZnP)_2$ | 0.001057 |
| mp-1820 | YbS | 0.000485 | mp-11824 | $Ca_3PN$ | 0.001057 |
| mp-23949 | $RbCaH_3$ | 0.000488 | mp-8398 | $YbCsF_3$ | 0.001078 |
| mp-1672 | CaS | 0.000524 | mp-1186 | $ZrS_2$ | 0.001097 |
| mp-644203 | $CsCaH_3$ | 0.000524 | mp-10401 | $Tl_2SnF_6$ | 0.001097 |
| mp-470 | $GeO_2$ | 0.000525 | mp-1039 | MgTe | 0.001118 |
| mp-13031 | MgSe | 0.000526 | mp-380 | ZnSe | 0.001128 |
| mp-10695 | ZnS | 0.000531 | mp-10933 | $Nd_2TeS_2$ | 0.001129 |
| mp-1315 | MgS | 0.000535 | mp-19921 | PbO | 0.001129 |
| mp-1986 | ZnO | 0.000535 | mp-23193 | KCl | 0.001137 |
| mp-9514 | $Mg(AlC)_2$ | 0.000545 | mp-570259 | $MgCl_2$ | 0.001138 |
| mp-149 | Si | 0.000559 | mp-632319 | CsH | 0.001156 |
| mp-8192 | $Rb_2PtF_6$ | 0.000576 | mp-9564 | $Ca(MgAs)_2$ | 0.001162 |
| mp-24721 | RbH | 0.000576 | mp-3654 | $RbCaF_3$ | 0.001168 |
| mp-8279 | $Ba(CdP)_2$ | 0.000578 | mp-14099 | $ZnAgF_3$ | 0.001168 |
| mp-1550 | AlP | 0.000579 | mp-4950 | $KCaF_3$ | 0.001172 |
| mp-4495 | KLiTe | 0.000579 | mp-774712 | LiCuS | 0.001175 |
| mp-2469 | CdS | 0.000581 | mp-8751 | RbLiS | 0.001177 |
| mp-856 | $SnO_2$ | 0.000582 | mp-7483 | $KHgF_3$ | 0.001179 |
| mp-2251 | $Li_3N$ | 0.000585 | mp-7738 | LaSeF | 0.001182 |
| mp-24423 | SrHBr | 0.000588 | mp-6948 | KNaO | 0.001191 |
| mp-553875 | $Zr_2SN_2$ | 0.000603 | mp-21855 | $VCu_3Se_4$ | 0.001195 |
| mp-8181 | $Li_2CeN_2$ | 0.000613 | mp-8399 | $CsCdF_3$ | 0.001206 |
| mp-7768 | $Th_2SeN_2$ | 0.000619 | mp-7104 | $CsCaF_3$ | 0.001225 |
| mp-1216 | YbO | 0.000622 | mp-12620 | NdSeF | 0.001234 |
| mp-9915 | LiBeP | 0.000623 | mp-6952 | $YbRbF_3$ | 0.001244 |
| mp-5795 | $CaMg_2N_2$ | 0.000627 | mp-1087 | SrS | 0.001248 |
| mp-8039 | $AlF_3$ | 0.000630 | mp-4824 | NaZnP | 0.001253 |
| mp-14254 | $NdAlO_3$ | 0.000636 | mp-9570 | $Ca(CdP)_2$ | 0.001253 |
| mp-10616 | BaLiAs | 0.000637 | mp-7891 | $Mg_3As_2$ | 0.001269 |
| mp-7604 | $Mg_3NF_3$ | 0.000641 | mp-9567 | $Ba(MgSb)_2$ | 0.001272 |
| mp-2691 | CdSe | 0.000645 | mp-29342 | $Ca_3PCl_3$ | 0.001272 |
| mp-28065 | ThNCl | 0.000646 | mp-10760 | MgSe | 0.001293 |
| mp-2657 | $TiO_2$ | 0.000656 | mp-7786 | $CsCu_3S_2$ | 0.001305 |
| mp-753920 | $Tm_2SeO_2$ | 0.000668 | mp-1253 | BaSe | 0.001306 |
| mp-2574 | $ZrO_2$ | 0.000673 | mp-8453 | RbNaO | 0.001322 |
| mp-9582 | $Yb(ZnP)_2$ | 0.000675 | mp-4043 | $NbCu_3Se_4$ | 0.001326 |
| mp-9517 | $SrTiN_2$ | 0.000678 | mp-1500 | BaS | 0.001341 |
| mp-13032 | MgS | 0.000681 | mp-22914 | CuCl | 0.001352 |
| mp-1249 | $MgF_2$ | 0.000685 | mp-8280 | $Ba(MgAs)_2$ | 0.001377 |
| mp-9250 | RbLiSe | 0.000694 | mp-5962 | NaMgAs | 0.001378 |
| mp-3614 | $KTaO_3$ | 0.000704 | mp-406 | CdTe | 0.001380 |
| mp-8191 | $Cs_2PtF_6$ | 0.000704 | mp-6951 | $RbCdF_3$ | 0.001396 |
| mp-755340 | $Tb_2SeO_2$ | 0.000704 | mp-9566 | $Sr(MgSb)_2$ | 0.001397 |
| mp-24424 | BaHBr | 0.000706 | mp-8397 | $CsSrF_3$ | 0.001420 |
| mp-10550 | $SrMg_2N_2$ | 0.000709 | mp-28069 | $Ca_3AsCl_3$ | 0.001425 |
| mp-8231 | ZrSO | 0.000716 | mp-10175 | $KCdF_3$ | 0.001428 |
| mp-9575 | LiBeSb | 0.000731 | mp-2341 | $Li_3N$ | 0.001453 |

| id | Name | $P_3$ (cm) | id | Name | $P_3$ (cm) |
|---|---|---|---|---|---|
| mp-3821 | $K_2PtF_6$ | 0.000733 | mp-4081 | $TaCu_3Se_4$ | 0.001475 |
| mp-4675 | $NaTaO_3$ | 0.000734 | mp-9565 | $Ca(MgSb)_2$ | 0.001499 |
| mp-9912 | $Li_2CeP_2$ | 0.000739 | mp-23259 | LiBr | 0.001512 |
| mp-7949 | $Rb_2GeF_6$ | 0.000741 | mp-21043 | $RbPbF_3$ | 0.001519 |
| mp-7979 | $K_2PdF_6$ | 0.000744 | mp-1342 | BaO | 0.001523 |
| mp-4170 | $NaTaO_3$ | 0.000746 | mp-7090 | NaMgSb | 0.001548 |
| mp-7950 | ThSeO | 0.000746 | mp-7089 | KMgSb | 0.001558 |
| mp-4342 | $KNbO_3$ | 0.000747 | mp-1958 | SrTe | 0.001575 |
| mp-10086 | YSF | 0.000747 | mp-22894 | AgI | 0.001576 |
| mp-3136 | $NaNbO_3$ | 0.000752 | mp-665 | $SnSe_2$ | 0.001621 |
| mp-3970 | $K_2TiF_6$ | 0.000758 | mp-7548 | $BaSe_3$ | 0.001631 |
| mp-5229 | $SrTiO_3$ | 0.000765 | mp-8799 | RbNaS | 0.001656 |
| mp-27984 | PrClO | 0.000765 | mp-1519 | CaTe | 0.001659 |
| mp-20459 | $TiPbO_3$ | 0.000782 | mp-9295 | $TaCu_3Te_4$ | 0.001678 |
| mp-27823 | SmClO | 0.000784 | mp-5339 | CsNaTe | 0.001720 |
| mp-10322 | $BaHfN_2$ | 0.000787 | mp-11718 | RbF | 0.001729 |
| mp-3104 | $BaZrN_2$ | 0.000788 | mp-286 | YbSe | 0.001734 |
| mp-4419 | $NaNbO_3$ | 0.000797 | mp-6973 | CsNaS | 0.001743 |
| mp-5347 | $KAlF_4$ | 0.000798 | mp-23302 | RbI | 0.001749 |
| mp-1070 | CdSe | 0.000799 | mp-12908 | $ScAgSe_2$ | 0.001781 |
| mp-10694 | $ScF_3$ | 0.000800 | mp-28171 | $K_3IO$ | 0.001790 |
| mp-10748 | $TaCu_3S_4$ | 0.000803 | mp-27294 | $Ca_3AsBr_3$ | 0.001808 |
| mp-4764 | $Pr_2SeO_2$ | 0.000805 | mp-1000 | BaTe | 0.001837 |
| mp-10219 | YOF | 0.000812 | mp-22916 | NaBr | 0.001844 |
| mp-752658 | $Y_2SeO_2$ | 0.000812 | mp-22917 | CuBr | 0.001847 |
| mp-4551 | $SrHfO_3$ | 0.000819 | mp-561947 | $CsHgF_3$ | 0.001856 |
| mp-7825 | $K_2GeF_6$ | 0.000819 | mp-5811 | $CsPbF_3$ | 0.001869 |
| mp-8277 | $Sr(CdP)_2$ | 0.000819 | mp-1779 | YbTe | 0.001872 |
| mp-5986 | $BaTiO_3$ | 0.000821 | mp-7482 | $RbHgF_3$ | 0.001883 |
| mp-9486 | $K_2AlF_5$ | 0.000821 | mp-23268 | NaI | 0.001904 |
| mp-12992 | $BaTiO_3$ | 0.000821 | mp-9846 | RbCaSb | 0.001942 |
| mp-13033 | MgTe | 0.000825 | mp-27138 | $K_2PdBr_4$ | 0.001956 |
| mp-27985 | YClO | 0.000825 | mp-7434 | NaCuTe | 0.001968 |
| mp-2998 | $BaTiO_3$ | 0.000827 | mp-9200 | $K_3AuO$ | 0.001983 |
| mp-8452 | NaLiS | 0.000828 | mp-9845 | RbCaAs | 0.002041 |
| mp-10733 | $Sm_2O_3$ | 0.000828 | mp-9385 | $RbAu_3Se_2$ | 0.002043 |
| mp-12673 | $Lu_2SO_2$ | 0.000833 | mp-9386 | $CsAu_3Se_2$ | 0.002084 |
| mp-3556 | $Tm_2SO_2$ | 0.000834 | mp-23295 | RbCl | 0.002095 |
| mp-12894 | $Y_2SO_2$ | 0.000834 | mp-28650 | $CsBr_2F$ | 0.002095 |
| mp-28593 | $Li_3BrO$ | 0.000836 | mp-9384 | $CsAu_3S_2$ | 0.002100 |
| mp-3834 | $BaZrO_3$ | 0.000842 | mp-568544 | $CsCdCl_3$ | 0.002107 |
| mp-12671 | $Er_2SO_2$ | 0.000845 | mp-23251 | KBr | 0.002119 |
| mp-19845 | $TiPbO_3$ | 0.000849 | mp-22867 | RbBr | 0.002131 |
| mp-2763 | $Nd_2O_3$ | 0.000853 | mp-8658 | CsNaSe | 0.002137 |
| mp-3448 | $KMgF_3$ | 0.000854 | mp-1784 | CsF | 0.002146 |
| mp-5606 | $AlTlF_4$ | 0.000854 | mp-27243 | $K_2PtBr_4$ | 0.002184 |
| mp-12670 | $Ho_2SO_2$ | 0.000854 | mp-12953 | $TmAgTe_2$ | 0.002197 |
| mp-4823 | $Na_2PdC_2$ | 0.000855 | mp-569346 | CuI | 0.002252 |

| id | Name | $P_3$ (cm) | id | Name | $P_3$ (cm) |
|---|---|---|---|---|---|
| mp-9562 | LiBeAs | 0.000858 | mp-12902 | $ErAgTe_2$ | 0.002295 |
| mp-7769 | $Th_2SN_2$ | 0.000860 | mp-19717 | TePb | 0.002315 |
| mp-12669 | $Dy_2SO_2$ | 0.000861 | mp-12904 | $HoAgTe_2$ | 0.002325 |
| mp-2063 | $Pr_2O_3$ | 0.000863 | mp-12903 | $YAgTe_2$ | 0.002340 |
| mp-5621 | $NbCu_3S_4$ | 0.000864 | mp-22922 | AgCl | 0.002356 |
| mp-10919 | $Rb_2PtC_2$ | 0.000865 | mp-4024 | $DyAgTe_2$ | 0.002358 |
| mp-12668 | $Tb_2SO_2$ | 0.000870 | mp-30056 | $CsCaBr_3$ | 0.002358 |
| mp-8976 | $Cu_2WS_4$ | 0.000871 | mp-3551 | $TbAgTe_2$ | 0.002392 |
| mp-567290 | LaN | 0.000873 | mp-12779 | CdTe | 0.002428 |
| mp-22862 | NaCl | 0.000874 | mp-8234 | $BaTe_3$ | 0.002456 |
| mp-10918 | $Rb_2PdC_2$ | 0.000875 | mp-22906 | CsBr | 0.002508 |
| mp-8276 | $Sr(ZnP)_2$ | 0.000885 | mp-22898 | KI | 0.002512 |
| mp-13947 | $Rb_2HfF_6$ | 0.000888 | mp-573697 | CsCl | 0.002642 |
| mp-13946 | $Rb_2ZrF_6$ | 0.000897 | mp-23037 | $CsPbCl_3$ | 0.002682 |
| mp-7903 | $Cs_2ZrF_6$ | 0.000902 | mp-570418 | $YbI_2$ | 0.002780 |
| mp-22925 | AgI | 0.000904 | mp-567259 | $CdI_2$ | 0.002822 |
| mp-13948 | $Cs_2HfF_6$ | 0.000905 | mp-23231 | AgBr | 0.002842 |
| mp-5598 | $Sm_2SO_2$ | 0.000909 | mp-22865 | CsCl | 0.002936 |
| mp-8152 | $Li_2CeAs_2$ | 0.000913 | mp-570231 | $CsCdBr_3$ | 0.003025 |
| mp-755309 | $Li_3NbS_4$ | 0.000922 | mp-22903 | RbI | 0.003090 |
| mp-1968 | $La_2O_3$ | 0.000925 | mp-570223 | $CsGeBr_3$ | 0.003149 |
| mp-2758 | SrSe | 0.000925 | mp-27214 | $CsSnBr_3$ | 0.003177 |
| mp-3211 | $Nd_2SO_2$ | 0.000928 | mp-600089 | $CsPbBr_3$ | 0.003234 |
| mp-7297 | $Cs_2SnF_6$ | 0.000928 | mp-569639 | TlCl | 0.003304 |
| mp-22899 | LiI | 0.000930 | mp-23167 | TlCl | 0.003344 |
| mp-463 | KF | 0.000934 | mp-571222 | CsBr | 0.003408 |
| mp-560399 | $NaMgF_3$ | 0.000934 | mp-571458 | $RbGeI_3$ | 0.003521 |
| mp-10930 | TbSF | 0.000937 | mp-568560 | TlBr | 0.003797 |
| mp-3236 | $Pr_2SO_2$ | 0.000941 | mp-614603 | CsI | 0.004187 |
| mp-7100 | LaOF | 0.000941 | mp-22875 | TlBr | 0.004233 |
| mp-568273 | LiI | 0.000941 | mp-13548 | $Cs_2Pt$ | 0.004822 |
| mp-3931 | SmSF | 0.000949 | mp-571102 | TlI | 0.004915 |
| mp-22905 | LiCl | 0.000951 | mp-2667 | CsAu | 0.005085 |

**Table S2. Training data set of $\kappa_L$ for 45 materials.**

| id | Name | $\kappa_L$ (W/mK) | id | Name | $\kappa_L$ (W/mK) |
|---|---|---|---|---|---|
| mp-22922 | AgCl | 1 | mp-2176 | ZnTe | 18 |
| mp-22925 | AgI | 1.05 | mp-682 | NaF | 18.4 |
| mp-22914 | CuCl | 1.44 | mp-1190 | ZnSe | 19 |
| mp-23268 | NaI | 1.8 | mp-2605 | CaO | 27 |
| mp-2574 | ZrO2 | 1.9 | mp-10695 | ZnS | 27 |
| mp-1342 | BaO | 2.3 | mp-2624 | AlSb | 56 |
| mp-22903 | RbI | 2.3 | mp-1265 | MgO | 60 |
| mp-23302 | RbI | 2.3 | mp-1986 | ZnO | 60 |
| mp-22917 | CuBr | 2.52 | mp-252 | BeTe | 77.5 |
| mp-22898 | KI | 2.6 | mp-1550 | AlP | 90 |
| mp-463 | KF | 2.68 | mp-1541 | BeSe | 95 |
| mp-22916 | NaBr | 2.8 | mp-856 | $SnO_2$ | 98 |
| mp-23295 | RbCl | 2.8 | mp-2172 | AlAs | 98 |
| mp-23251 | KBr | 3.4 | mp-2490 | GaP | 100 |
| mp-22867 | RbBr | 3.8 | mp-422 | BeS | 143 |
| mp-2691 | CdSe | 4.4 | mp-149 | Si | 153 |
| mp-22862 | NaCl | 7.1 | mp-830 | GaN | 210 |
| mp-23193 | KCl | 7.1 | mp-661 | AlN | 285 |
| mp-406 | CdTe | 7.5 | mp-1479 | BP | 580 |
| mp-2657 | $TiO_2$ | 10 | mp-1700 | AlN | 350 |
| mp-2472 | SrO | 12 | mp-8062 | SiC | 479 |
| mp-2469 | CdS | 16 | mp-7599 | BeO | 370 |
| mp-1138 | LiF | 17.6 | | | |

**Table S3. Top-100 materials with lowest $P_3$.**

| Order | id | Name | Calculated $P_3$ ($10^{-4}$ cm) | Order | id | Name | Calculated $P_3$ ($10^{-4}$ cm) |
|---|---|---|---|---|---|---|---|
| 1 | mp-10044 | BAs | 0.6397 | 51 | mp-568028 | C | 2.7583 |
| 2 | mp-984718 | BAs | 0.9064 | 52 | mp-252 | BeTe | 2.7600 |
| 3 | mp-66 | C | 1.0005 | 53 | mp-604884 | BN | 2.7793 |
| 4 | mp-47 | C | 1.0335 | 54 | mp-629015 | BN | 2.7873 |
| 5 | mp-611426 | C | 1.2437 | 55 | mp-642462 | $B_3C_{10}N_3$ | 2.7900 |
| 6 | mp-616440 | C | 1.2569 | 56 | mp-13150 | BN | 2.7914 |
| 7 | mp-1569 | $Be_2C$ | 1.2596 | 57 | mp-984 | BN | 2.7989 |
| 8 | mp-1639 | BN | 1.3300 | 58 | mp-567885 | $C_3N_4$ | 2.8155 |
| 9 | mp-30148 | $BC_2N$ | 1.3670 | 59 | mp-7991 | BN | 2.8284 |
| 10 | mp-2653 | BN | 1.4394 | 60 | mp-28395 | $B_6P$ | 2.8292 |
| 11 | mp-2852 | $C_3N_4$ | 1.4490 | 61 | mp-570572 | $C_3N_4$ | 2.8906 |
| 12 | mp-571653 | $C_3N_4$ | 1.4529 | 62 | mp-1599 | BN | 2.9035 |
| 13 | mp-629458 | $BC_2N$ | 1.5127 | 63 | mp-971683 | $C_3N_4$ | 2.9110 |
| 14 | mp-15703 | $BeCN_2$ | 1.5472 | 64 | mp-344 | BN | 2.9309 |
| 15 | mp-611448 | C | 1.5482 | 65 | mp-7599 | BeO | 2.9800 |
| 16 | mp-569567 | C | 1.5531 | 66 | mp-8062 | SiC | 3.0200 |
| 17 | mp-569517 | C | 1.5547 | 67 | mp-563 | $C_3N_4$ | 3.0887 |
| 18 | mp-11276 | BeRh | 1.5567 | 68 | mp-11917 | $Mg(BeN)_2$ | 3.0900 |
| 19 | mp-971684 | $C_3N_4$ | 1.5763 | 69 | mp-645279 | $C_{68}OF_{20}$ | 3.0945 |
| 20 | mp-13151 | BN | 1.5954 | 70 | mp-683965 | $C_2F$ | 3.1442 |
| 21 | mp-717 | $B_2O_3$ | 1.6444 | 71 | mp-645316 | $C_7F_3$ | 3.1582 |
| 22 | mp-24 | C | 1.8006 | 72 | mp-989466 | $B_2(CN_2)_3$ | 3.1642 |
| 23 | mp-1541 | BeSe | 1.9400 | 73 | mp-569299 | $Be(BC)_2$ | 3.2416 |
| 24 | mp-18337 | $Be_3N_2$ | 2.0966 | 74 | mp-569655 | BN | 3.2452 |
| 25 | mp-9410 | $C_3N_4$ | 2.1532 | 75 | mp-989459 | $B_2(CN_2)_3$ | 3.3400 |
| 26 | mp-696746 | $B_4C$ | 2.1673 | 76 | mp-1479 | BP | 3.3600 |
| 27 | mp-160 | B | 2.1803 | 77 | mp-1700 | AlN | 3.3700 |
| 28 | mp-644751 | BN | 2.1808 | 78 | mp-1330 | AlN | 3.3969 |
| 29 | mp-632329 | C | 2.2215 | 79 | mp-306 | $B_2O_3$ | 3.4263 |
| 30 | mp-1346 | $B_6O$ | 2.2525 | 80 | mp-624 | $B_6As$ | 3.4475 |
| 31 | mp-990448 | C | 2.2559 | 81 | mp-6977 | $Be_3N_2$ | 3.4887 |
| 32 | mp-758933 | $B_8O$ | 2.2719 | 82 | mp-30935 | $B_{17}F_{27}$ | 3.5351 |
| 33 | mp-568806 | C | 2.2757 | 83 | mp-30142 | $Be_4N_6O_{19}$ | 3.5458 |
| 34 | mp-32715 | $B_{36}O_5$ | 2.3104 | 84 | mp-2172 | AlAs | 3.6100 |
| 35 | mp-561543 | $BeF_2$ | 2.3104 | 85 | mp-2624 | AlSb | 3.6400 |
| 36 | mp-569304 | C | 2.3132 | 86 | mp-30141 | $Be(N_3O_7)_2$ | 3.7489 |
| 37 | mp-644802 | $C_{17}F_5$ | 2.3167 | 87 | mp-6988 | FeN | 3.7733 |
| 38 | mp-830 | GaN | 2.3300 | 88 | mp-3589 | $BPO_4$ | 3.7926 |
| 39 | mp-1778 | BeO | 2.3796 | 89 | mp-11653 | $BPO_4$ | 3.7958 |
| 40 | mp-530033 | $B_8O$ | 2.4404 | 90 | mp-8881 | AlAs | 3.8100 |
| 41 | mp-630227 | C | 2.4462 | 91 | mp-554023 | $Be_2BO_3F$ | 3.8297 |

| Order | id | Name | Calculated $P_3$ ($10^{-4}$ cm) | Order | id | Name | Calculated $P_3$ ($10^{-4}$ cm) |
|---|---|---|---|---|---|---|---|
| 42 | mp-647169 | $C_{10}F_3$ | 2.4922 | 92 | mp-422 | BeS | 3.9700 |
| 43 | mp-2542 | BeO | 2.5200 | 93 | mp-975644 | $BeF_2$ | 3.9705 |
| 44 | mp-1985 | $C_3N_4$ | 2.5206 | 94 | mp-8756 | KLiSe | 3.9800 |
| 45 | mp-804 | GaN | 2.5730 | 95 | mp-4674 | $BNF_8$ | 4.0054 |
| 46 | mp-989468 | $B_2(CN_2)_3$ | 2.6186 | 96 | mp-570002 | C | 4.0385 |
| 47 | mp-1265 | MgO | 2.6200 | 97 | mp-617 | $PtO_2$ | 4.0400 |
| 48 | mp-568286 | C | 2.6432 | 98 | mp-555207 | $BN(OF_2)_2$ | 4.0916 |
| 49 | mp-680372 | C | 2.6659 | 99 | mp-2490 | GaP | 4.2500 |
| 50 | mp-683919 | C | 2.7542 | 100 | mp-30936 | $B_5F_6$ | 4.2531 |

**Table S4. Comparison of ordinary machine learning and transfer learning**. The predicted lattice thermal conductivity by ordinary machine learning and transfer learning for confirmed top-14 materials with small scattering phase space.

| mp-ID | Name | Ordinary machine learning ($Wm^{-1}K^{-1}$) | Transfer learning ($Wm^{-1}K^{-1}$) |
|---|---|---|---|
| 10044 | cubic BAs | 181 | 274 |
| 984718 | wurtzite BAs | 188 | 280 |
| 66 | diamond | 75 | 2168 |
| 47 | lonsdaleite | 352 | 2166 |
| 611426 | C | 363 | 2061 |
| 616440 | C | 375 | 1990 |
| 1569 | $Be_2C$ | 231 | 651 |
| 1639 | cubic BN | 431 | 1048 |
| 30148 | $BC_2N$ | 468 | 1314 |
| 2653 | wurtzite BN | 446 | 1102 |
| 2852 | cubic $C_3N_4$ | 638 | 2041 |
| 571653 | pseudo $C_3N_4$ | 619 | 2145 |
| 629458 | $BC_2N$ | 457 | 1293 |
| 15703 | $BeCN_2$ | 332 | 814 |

**Table S5. Descriptors relevant to $\kappa_L$ and irrelevant to $P_3$.** Descriptors with ($MIC_P$ - $MIC_\kappa$ < -0.09) were listed.

| Features | $MIC_P$ - $MIC_\kappa$ | $MIC_P$ | $MIC_\kappa$ |
|---|---|---|---|
| ave.c6_gb | -0.1540 | 0.2893 | 0.4432 |
| ave.Polarizability | -0.1465 | 0.2671 | 0.4136 |
| ave.dipole_polarizability | -0.1406 | 0.2623 | 0.4028 |
| max.vdw_radius_alvarez | -0.1393 | 0.2454 | 0.3847 |
| max.dipole_polarizability | -0.1298 | 0.2078 | 0.3375 |
| max.Polarizability | -0.1269 | 0.2114 | 0.3383 |
| min.mendeleev_number | -0.1258 | 0.2175 | 0.3433 |
| max.period | -0.1241 | 0.2475 | 0.3716 |
| var.c6_gb | -0.1229 | 0.2396 | 0.3626 |
| max.covalent_radius_cordero | -0.1208 | 0.2448 | 0.3656 |
| max.covalent_radius_pyykko | -0.1194 | 0.2460 | 0.3653 |
| max.vdw_radius | -0.1189 | 0.2527 | 0.3716 |
| var.Polarizability | -0.1186 | 0.1990 | 0.3175 |
| min.en_allen | -0.1176 | 0.1962 | 0.3138 |
| max.c6_gb | -0.1142 | 0.2500 | 0.3642 |
| sum.c6_gb | -0.1053 | 0.1621 | 0.2674 |
| var.dipole_polarizability | -0.1052 | 0.2012 | 0.3064 |
| min.first_ion_en | -0.1032 | 0.2151 | 0.3182 |
| max.covalent_radius_slater | -0.1016 | 0.2637 | 0.3653 |
| min.en_pauling | -0.0998 | 0.2012 | 0.3010 |
| min.electron_negativity | -0.0991 | 0.2046 | 0.3036 |
| max.atomic_radius_rahm | -0.0921 | 0.1759 | 0.2680 |
| max.vdw_radius_mm3 | -0.0917 | 0.2504 | 0.3421 |
| ave.mendeleev_number | -0.0907 | 0.2192 | 0.3098 |
| var.num_f_valence | -0.0903 | 0.2084 | 0.2987 |
| sum.num_f_valence | -0.0901 | 0.2130 | 0.3031 |

**Table S6. Descriptors relevant to $P_3$ and irrelevant to $\kappa_L$.** Descriptors with ($MIC_P$ - $MIC_\kappa$ > 0.09) were listed.

| Features | $MIC_P$ - $MIC_\kappa$ | $MIC_P$ | $MIC_\kappa$ |
|---|---|---|---|
| sum.specific_heat | 0.0905 | 0.2838 | 0.1933 |
| min.covalent_radius_pyykko_double | 0.0926 | 0.3138 | 0.2212 |
| min.vdw_radius_mm3 | 0.0934 | 0.3180 | 0.2246 |
| max.mendeleev_number | 0.0944 | 0.3163 | 0.2219 |
| min.Polarizability | 0.0950 | 0.3199 | 0.2249 |
| min.dipole_polarizability | 0.0958 | 0.3202 | 0.2244 |
| sum.bulk_modulus | 0.0965 | 0.2914 | 0.1949 |
| min.gs_est_bcc_latcnt | 0.0970 | 0.3660 | 0.2691 |
| min.gs_volume_per | 0.0979 | 0.3687 | 0.2708 |
| max.specific_heat | 0.0980 | 0.2856 | 0.1876 |
| ave.heat_capacity_mass | 0.0982 | 0.3856 | 0.2874 |
| min.atomic_radius_rahm | 0.0987 | 0.3231 | 0.2245 |
| min.covalent_radius_slater | 0.0990 | 0.3220 | 0.2230 |
| min.c6_gb | 0.0991 | 0.3232 | 0.2241 |
| min.gs_est_fcc_latcnt | 0.0995 | 0.3658 | 0.2663 |
| min.vdw_radius_alvarez | 0.1014 | 0.3283 | 0.2269 |
| min.boiling_point | 0.1015 | 0.3421 | 0.2406 |
| max.num_d_valence | 0.1035 | 0.2681 | 0.1646 |
| min.covalent_radius_cordero | 0.1038 | 0.3259 | 0.2221 |
| min.fusion_enthalpy | 0.1047 | 0.3388 | 0.2342 |
| min.melting_point | 0.1056 | 0.3336 | 0.2281 |
| min.covalent_radius_pyykko | 0.1074 | 0.3324 | 0.2250 |
| min.vdw_radius | 0.1083 | 0.3299 | 0.2216 |
| min.density | 0.1101 | 0.3114 | 0.2013 |
| min.hhi_r | 0.1117 | 0.3426 | 0.2309 |
| min.period | 0.1118 | 0.3729 | 0.2611 |
| min.evaporation_heat | 0.1162 | 0.3228 | 0.2066 |
| max.heat_capacity_mass | 0.1207 | 0.3143 | 0.1936 |
| min.atomic_number | 0.1251 | 0.3926 | 0.2676 |
| min.atomic_weight | 0.1259 | 0.3921 | 0.2662 |
| sum.heat_capacity_mass | 0.1270 | 0.3319 | 0.2050 |

**Table S7. Comparison of relaxation time approximation and iterative Boltzmann transport equation solution for top-14 materials**.

| mp-ID | Name | Structure | Thermal conductivity ($Wm^{-1}K^{-1}$) | | | | | |
|---|---|---|---|---|---|---|---|---|
| | | | relaxation time approximation | | | iterative solution | | |
| | | | xx | yy | zz | xx | yy | zz |
| 10044 | cubic BAs | F-43m | 1403 | 1403 | 1403 | 3411 | 3411 | 3411 |
| 984718 | wurtzite BAs | $P6_3mc$ | 1700 | 1700 | 1650 | 2947 | 2947 | 1881 |
| 66 | diamond | Fd-3m | 1957 | 1957 | 1957 | 3048 | 3048 | 3048 |
| 47 | lonsdaleite | $P6_3/mmc$ | 1873 | 1873 | 1753 | 2533 | 2533 | 2122 |
| 611426 | C | $P6_3/mmc$ | 2165 | 2165 | 2411 | 2842 | 2842 | 2675 |
| 616440 | C | $P6_3/mmc$ | 1921 | 1921 | 3834 | 2583 | 2583 | 4214 |
| 1569 | $Be_2C$ | Fm-3m | 112 | 112 | 112 | 117 | 117 | 117 |
| 1639 | cubic BN | F-43m | 1219 | 1219 | 1219 | 1876 | 1876 | 1876 |
| 30148 | $BC_2N$ | $P222_1$ | 830 | 844 | 739 | 895 | 910 | 804 |
| 2653 | wurtzite BN | $P6_3mc$ | 981 | 981 | 1035 | 1359 | 1359 | 1305 |
| 2852 | cubic $C_3N_4$ | I-43d | 229 | 229 | 229 | 234 | 234 | 234 |
| 571653 | pseudo $C_3N_4$ | P-43m | 266 | 266 | 266 | 275 | 275 | 275 |
| 629458 | $BC_2N$ | Pmm2 | 1102 | 755 | 702 | 1392 | 972 | 784 |
| 15703 | $BeCN_2$ | I-42d | 327 | 327 | 412 | 351 | 351 | 440 |

**Table S8. Parameter settings for ShengBTE calculation**.

| mp-ID | Name | Number of atoms in unit cell | DFPT q-point grid | Cell size for cubic force constants | Neighbor cutoff for cubic force constants | *q*-mesh size |
|---|---|---|---|---|---|---|
| 10044 | cubic BAs | 2 | 8×8×8 | 4×4×4 | 5 | 22×22×22 |
| 984718 | wurtzite BAs | 4 | 4×4×4 | 4×4×4 | 3 | 22×22×22 |
| 66 | diamond | 2 | 6×6×6 | 4×4×4 | 5 | 21×21×21 |
| 47 | lonsdaleite | 4 | 4×4×4 | 4×4×4 | 5 | 20×20×20 |
| 611426 | C | 8 | 4×4×4 | 4×4×4 | 3 | 12×12×12 |
| 616440 | C | 16 | 4×4×4 | 3×3×2 | 3 | 7×7×7 |
| 1569 | $Be_2C$ | 3 | 6×6×6 | 4×4×4 | 3 | 16×16×16 |
| 1639 | cubic BN | 2 | 8×8×8 | 4×4×4 | 5 | 22×22×22 |
| 30148 | $BC_2N$ | 8 | 3×3×3 | 3×3×3 | 3 | 12×12×12 |
| 2653 | wurtzite BN | 4 | 4×4×4 | 4×4×2 | 5 | 15×15×15 |
| 2852 | cubic $C_3N_4$ | 28 | 2×2×2 | 2×2×2 | 3 | 7×7×7 |
| 571653 | pseudo $C_3N_4$ | 7 | 4×4×4 | 3×3×3 | 5 | 20×20×20 |
| 629458 | $BC_2N$ | 4 | 4×4×4 | 4×4×4 | 5 | 15×15×15 |
| 15703 | $BeCN_2$ | 16 | 2×2×2 | 2×2×2 | 3 | 7×7×7 |

**Note A. Phonon Boltzmann transport equation**

In order to solve phonon Boltzmann transport equation, the scattering rate *P* of three-phonon scattering process is calculated according to Fermi's golden rule,

$$P_i^f = \frac{2\pi}{\hbar}\left|\langle f \mid V_3 | i \rangle\right|^2 \delta(E_f - E_i), \tag{S1}$$

where *i* and *f* indicate initial and final states, and $V_3$ is the three phonon coupling potential. The three-phonon scattering evens can be divided into two types as shown in **figure S1**, type (a): phonon mode (***qs***) absorbs mode (***q's'***) and generate the third phonon mode (***q"s"***), and type (b): phonon mode (***qs***) decomposes into two phonon modes (***q's'***) and (***q"s"***). Both of the processes should satisfy momentum and energy conservation.

The net scattering rate involving a type (a) event is given by,

$$P_{qs,q's'}^{q"s"} - P_{q"s"}^{qs,q's'} = \tilde{P}_{qs,q's'}^{q"s"}\left(\psi_{qs} + \psi_{q's'} - \psi_{q"s"}\right), \tag{S2}$$

where, $\psi$ is the first order perturbation of phonon population *n*,

$$\tilde{P}_{qs,q's'}^{q"s"} = 2\pi \bar{n}_{qs}\bar{n}_{q's'}(\bar{n}_{q"s"}+1)\left|\tilde{V}_3(-\boldsymbol{q}s,-\boldsymbol{q}'s',\boldsymbol{q}"s")\right|^2 \delta(\omega_{qs}+\omega_{q's'}-\omega_{q"s"}). \tag{S3}$$

Similarly, the net scattering rate involving a type (b) event is given by,

$$P_{qs}^{q's',q"s"} - P_{q's',q"s"}^{qs} = \tilde{P}_{qs}^{q's',q"s"}\left(\psi_{qs} - \psi_{q's'} - \psi_{q"s"}\right), \tag{S4}$$

where,

$$\tilde{P}_{qs}^{q's',q"s"} = 2\pi \bar{n}_{qs}(\bar{n}_{q's'}+1)(\bar{n}_{q"s"}+1)\left|\tilde{V}_3(-\boldsymbol{q}s,\boldsymbol{q}'s',\boldsymbol{q}"s")\right|^2 \delta(\omega_{qs}-\omega_{q's'}-\omega_{q"s"}). \tag{S5}$$

The net scattering rate is the sum of scattering rates due to type (a) and (b), and the phonon Boltzmann transport equation can be rewritten as,

$$-v_{qs}\nabla T\frac{\partial \bar{n}_{qs}}{\partial T} = \sum_{q's',q"s"}\left[\tilde{P}_{qs,q's'}^{q"s"}(\psi_q^s + \psi_{q'}^{s'} - \psi_{q"}^{s"}) + \frac{1}{2}\tilde{P}_{qs}^{q's',q"s"}(\psi_q^s - \psi_{q'}^{s'} - \psi_{q"}^{s"})\right]. \tag{S6}$$

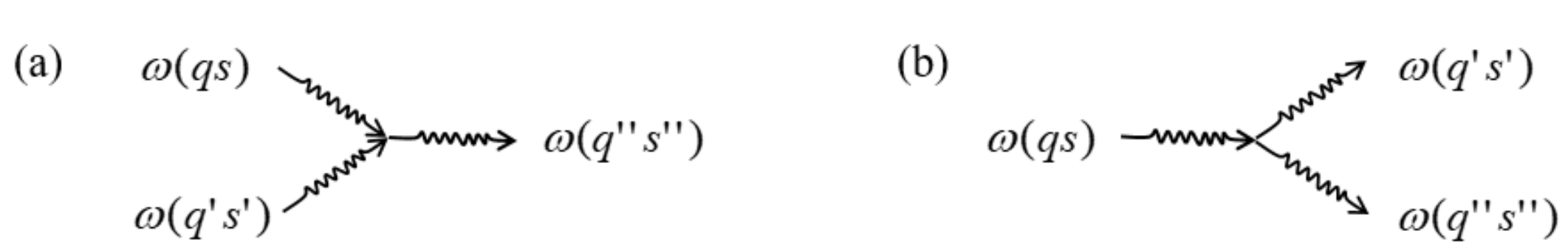

**Fig. S1**. **Types of three-phonon scattering processes**.

**Note B. Relaxation time approximation**

Under relaxation time approximation, $\psi_{q'}^{s'}, \psi_{q''}^{s''}=0$, the phonon relaxation time is given by,

$$\frac{1}{\tau_{qs}} = \pi \sum_{q's',s''} \left| \tilde{V}_3(-\boldsymbol{q}s, \boldsymbol{q}'s', \boldsymbol{q}''s'') \right|^2 \times \begin{bmatrix} 2(\bar{n}_{q's'} - \bar{n}_{q''s''})\delta(\omega_{qs} + \omega_{q's'} - \omega_{q''s''}) \\ +(1 + \bar{n}_{q's'} + \bar{n}_{q''s''})\delta(\omega_{qs} - \omega_{q's'} - \omega_{q''s''}) \end{bmatrix}, \tag{S7}$$

and the lattice thermal conductivity given by,

$$\kappa_{\alpha\beta} = \frac{\hbar^2}{N_q \Omega k_B T^2} \sum_{qs} c_{\alpha,qs} c_{\beta,qs} \omega_{qs}^{\ 2} \bar{n}_{qs} (\bar{n}_{qs} + 1) \tau_{qs}, \tag{S8}$$

where $c$ is group velocity, $\omega$ is frequency, $\alpha$ and $\beta$ indicate different lattice directions, $\Omega$ is the volume of the unit cell.

**Note C. Iterative solution of Boltzmann transport equation**

The relaxation time approximation solution of Boltzmann transport equation usually underestimates the thermal conductivity when dealing with high conductivity crystals (see Figure S2). In this situation, we can calculate the exact conductivity via iterative solution. By defining $\Psi_{qs} = \Sigma_\alpha F_{\alpha,qs}(\partial T / \partial x_\alpha)$, the Eq. S6 can be rewritten as,

$$-v_{qs} \frac{\hbar\omega_{qs}\bar{n}_{qs}(\bar{n}_{qs}+1)}{k_B T^2} = \sum_{q's',q''s''} \left[ \tilde{P}_{qs,q's'}^{q''s''}(F_{qs} + F_{q's'} - F_{q''s''}) + \frac{1}{2}\tilde{P}_{qs}^{q's',q''s''}(F_{qs} - F_{q's'} - F_{q''s''}) \right]. \tag{S9}$$

By defining $Q_{qs} = \sum_{q's',q''s''} \left[ \tilde{P}_{qs,q's'}^{q''s''} + \frac{1}{2}\tilde{P}_{qs}^{q's',q''s''} \right]$ and $-v_{qs} \frac{\hbar\omega_{qs}\bar{n}_{qs}(\bar{n}_{qs}+1)}{k_B T^2} = F_{\alpha,qs}^0 Q_{qs}$, we will have,

$$F_{\alpha,qs} = F^0_{\alpha,qs} + \frac{1}{Q_{qs}} \sum_{q's',q''s''} \left[ \tilde{P}^{q''s''}_{qs,q's'} (F_{\alpha,q''s''} - F_{\alpha,q's'}) + \frac{1}{2} \tilde{P}^{q's',q''s''}_{qs} (F_{\alpha,q's'} + F_{\alpha,q''s''}) \right]. \tag{S10}$$

At the beginning of iterative solution, the second term in Eq. S10 is zero, for the next iteration, $F_{\alpha,q's'}$ and $F_{\alpha,q''s''}$ are calculated according to the last step result. After several steps of iteration, convergence will be achieved for $F_{\alpha,qs}$, and the exact lattice thermal conductivity using iterative solution is given by,

$$\kappa_{\alpha\beta} = \frac{1}{N_0 \Omega k_B T^2} \sum_{qs} \left( \hbar \omega_{qs} \right)^2 \bar{n}_{qs} (\bar{n}_{qs} + 1) v_{\alpha,qs} F_{\beta,qs} . \tag{S11}$$

**Note D. Density-functional theory calculation details**

Density-functional theory calculations were performed using Quantum ESPRESSO [1] with the revised Perdew-Burke-Ernzerhof exchange-correlation functions based on the generalized gradient approximation, which improves the equilibrium properties for solids [2]. We employed kinetic energy cutoffs of 80 and 400 Ry for wave functions and charge density, respectively. The resolution of k-point mesh was set around 0.2/Å.

The thermal conductivity calculations were based on the harmonic and anharmonic cubic interatomic force constants. The lattice parameters and atomic positions for final thermal conductivity calculation were optimized until an energy convergence threshold of $10^{-8}$ eV and Hellmann-Feynman forces less than 2 meV/Å were achieved. The harmonic interatomic force constants were obtained via density functional perturbation theory, as implemented in Quantum ESPRESSO. The three-phonon scattering phase space was calculated by the ALAMODE package [3], where the q-mesh density was adjusted appropriately on the basis of the number of atoms in a primitive unit cell: 20×20×20 for atom numbers less than 5, 5×5×5 for atom numbers more than 20, and 10×10×10 for anything in between. The anharmonic cubic interatomic force constants were computed by the thirdorder.py tool provided in ShengBTE [4], which resolves an irreducible set of atomic displacements to compute the third-order interatomic force constants matrices. The detailed supercell size and neighbor cutoff settings for the ShengBTE calculation are listed in the Supplementary Information (see Supplementary **Table S8** for details).

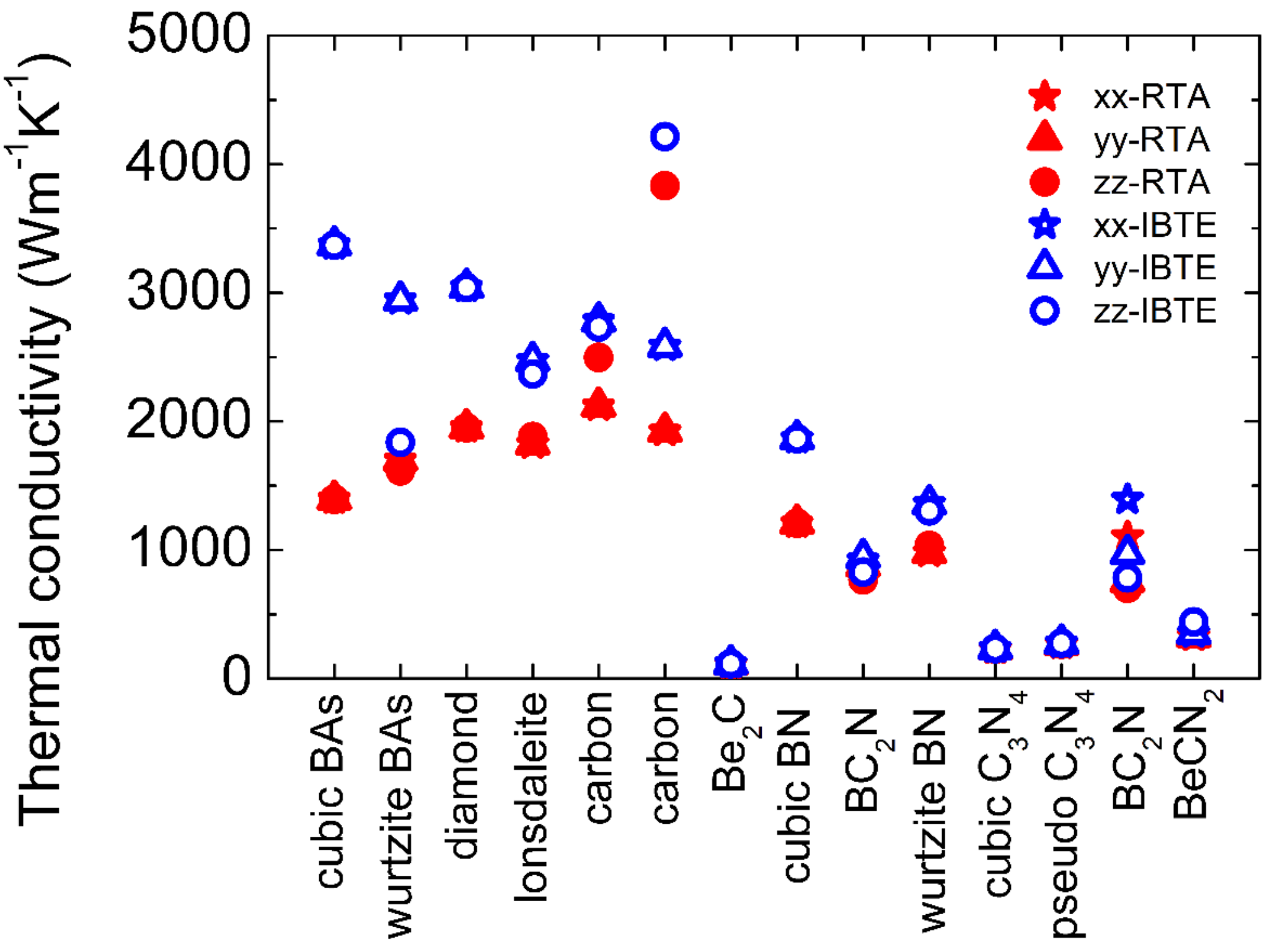

**Figure S2**. **Comparison of relaxation time approximation and iterative Boltzmann transport equation solution**. It is obvious that the thermal conductivity difference between relaxation time approximation (indicated as RTA in the figure) and iterative Boltzmann transport equation solution (indicated as IBTE in the figure) results is significantly large when thermal conductivity is higher than 1000 $Wm^{-1}K^{-1}$.

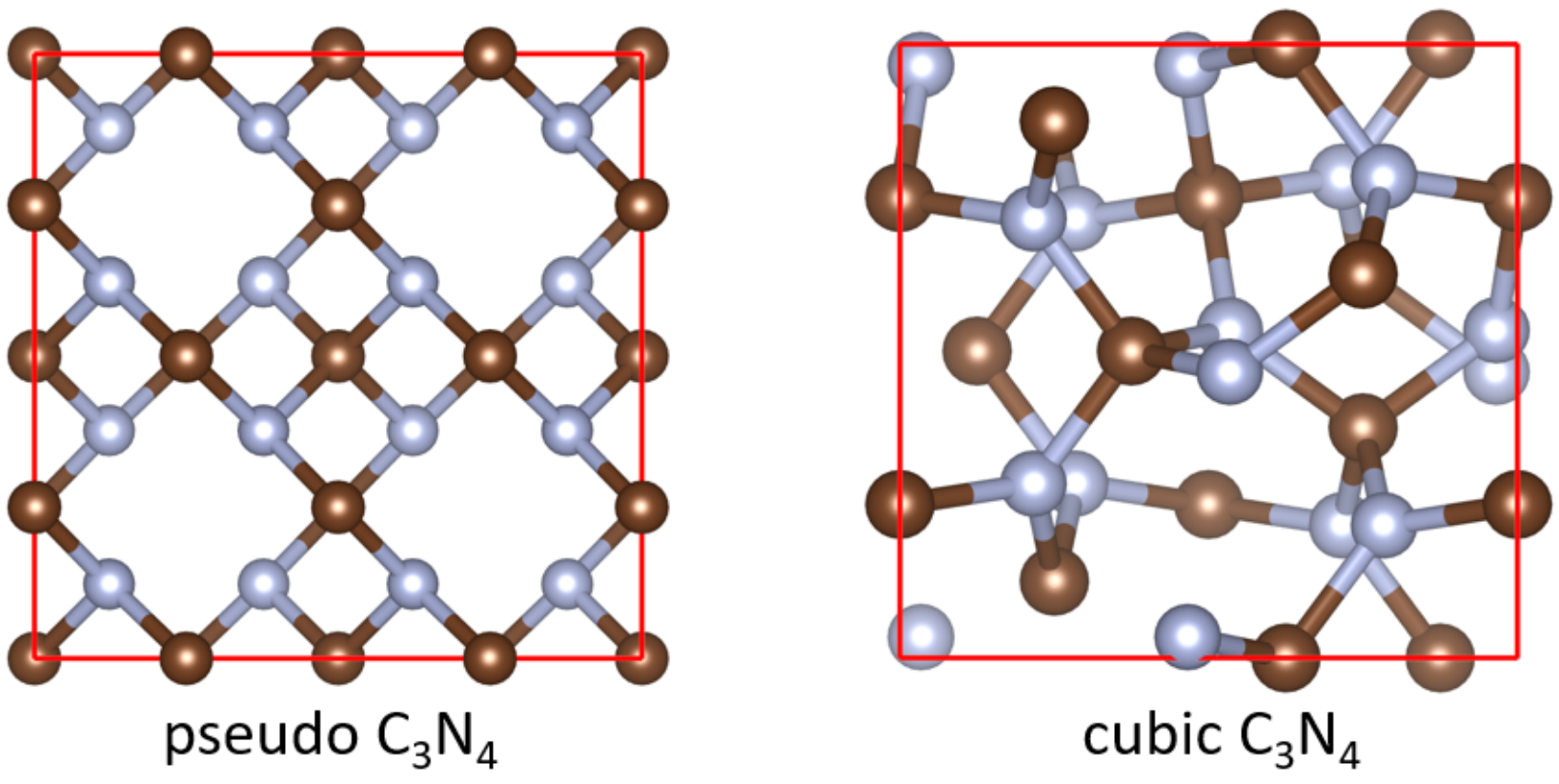

**Figure S3**. **The looks-defective and complex structures of pseudo and cubic $C_3N_4$**.

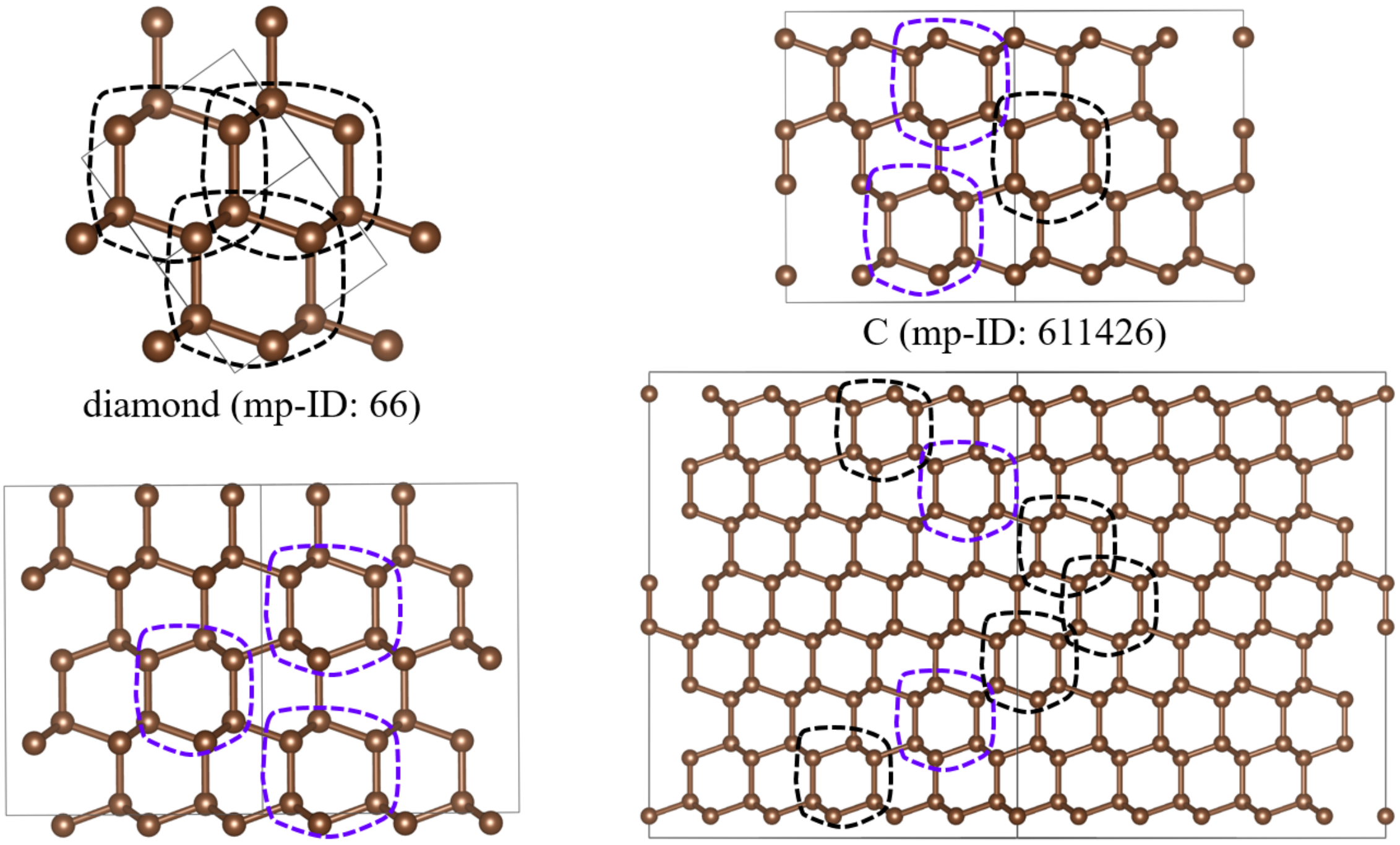

**Figure S4**. **Comparison of carbon allotropes**. Structures of diamond (mp-ID: 66), lonsdaleite (mp-ID: 47), C (mp-ID: 611426) and C (mp-ID: 616440) were shown. Black and blue circles indicate the typical C-rings of diamond and lonsdaleite, respectively.

## References

1. Giannozzi, P.; Baroni, S.; Bonini, N.; Calandra, M.; Car, R.; Cavazzoni, C.; Ceresoli, D.; Chiarotti, G. L.; Cococcioni, M.; Dabo, I.; Dal Corso, A.; de Gironcoli, S.; Fabris, S.; Fratesi, G.; Gebauer, R.; Gerstmann, U.; Gougoussis, C.; Kokalj, A.; Lazzeri, M.; Martin-Samos, L.; Marzari, N.; Mauri, F.; Mazzarello, R.; Paolini, S.; Pasquarello, A.; Paulatto, L.; Sbraccia, C.; Scandolo, S.; Sclauzero, G.; Seitsonen, A. P.; Smogunov, A.; Umari, P.; Wentzcovitch, R. M., QUANTUM ESPRESSO: a modular and open-source software project for quantum simulations of materials. *J Phys Condens Matter* **2009,** 21, (39), 395502.
2. Perdew, J. P.; Ruzsinszky, A.; Csonka, G. I.; Vydrov, O. A.; Scuseria, G. E.; Constantin, L. A.; Zhou, X.; Burke, K., Restoring the density-gradient expansion for exchange in solids and surfaces. *Phys Rev Lett* **2008,** 100, (13), 136406.
3. Tadano, T.; Gohda, Y.; Tsuneyuki, S., Anharmonic force constants extracted from first-principles molecular dynamics: applications to heat transfer simulations. *J Phys Condens Matter* **2014,** 26, (22), 225402.
4. Li, W.; Carrete, J.; A. Katcho, N.; Mingo, N., ShengBTE: A solver of the Boltzmann transport equation for phonons. *Computer Physics Communications* **2014,** 185, (6), 1747-1758.